\documentclass[twocolumn,showpacs,amsmath,
amssymb,nofootinbib]{revtex4-1}
\usepackage{dcolumn}
\usepackage{lipsum}
\usepackage{bm}
\usepackage{graphicx}
\usepackage{color}
\usepackage{amsmath}
\newcommand{\be}{\begin{equation}}
\newcommand{\ee}{\end{equation}}
\newcommand{\ba}{\begin{eqnarray}}
\newcommand{\ea}{\end{eqnarray}}

\newcommand{\n}[1]{\label{#1}}
\newcommand{\eq}[1]{Eq.(\ref{#1})}

\begin{document}

\title{Distorted Local Shadows}
\author{Shohreh Abdolrahimi ${}^{a,} \, {}^b$,}
\email{abdolrah@ualberta.ca}
\author{Robert B. Mann${}^{c}$}
\email{rbmann@uwaterloo.ca}
\author{and Christos Tzounis ${}^{a,} \, {}^b$}
\email{tzounis@ualberta.ca}
\affiliation{${}^a$ Theoretical Physics Institute, University of Alberta, Edmonton, AB, Canada,  T6G 2G7}
\affiliation{${}^b$ Institut f{\"u}r Physik, Universit{\"a}t Oldenburg, Postfach 2503 D-26111 Oldenburg, Germany}
\affiliation{${}^c$ Department of Physics and Astronomy, University of Waterloo, Waterloo, Ontario, Canada, N2L 3G1}
\date{\today}

\begin{abstract}
We introduce the notion of a local shadow for a black hole and determine its shape for the particular case of a distorted Schwarzschild black hole. Considering the lowest-order  even and odd multiple moments, we compute the relation between the deformations of the shadow of a Schwarzschild black hole and the distortion multiple moments. For the range of values of multiple moments that we consider, the horizon is deformed much less than its corresponding shadow, suggesting the horizon is more `rigid'.  Quite unexpectedly we find that a prolate distortion of the horizon gives rise to an oblate distortion of the shadow, and vice-versa.
\end{abstract} 

\pacs{04.70.-s, 04.70.Bw, 04.25.-g}
\maketitle



\section{Introduction}
The mysterious radio source, called Sagittarius $A^{*}$, at the centre of Milky Way, with a mass of around $4\times 10^6 M_{\odot}$ is considered to be a supermassive black hole. To a distant observer, the black hole casts a relatively large ``shadow'' with an apparent diameter of $\sim 10$ gravitational radii\footnote{The diameter of the ``shadow'' will be slightly different in the case of a rotating black hole.} due to bending of light by the black hole. The predicted size of the shadow of Sagittarius $A^{*}$, ($\sim 30\mu$arcsec),  will be probed in the next few decades, using the Event Horizon Telescope via Very Long Baseline Interferometry \cite{1}. 

The original investigation of the shadow of a Schwarzschild black hole \cite{Synge} has been extended to 
a number of other cases, including Kerr black holes \cite{11}, (see also \cite{19,20,20b}), Kerr-Newman black holes \cite{12}, Schwarzschild-de Sitter black holes \cite{12B},   Tomimatsu-Sato space-times \cite{13},  black holes in extended Chern-Simons modified gravity \cite{14},  rotating braneworld black holes \cite{15A,15},
 Kaluza-Klein rotating dilaton black holes \cite{16}, for the Kerr-NUT black holes \cite{17},   Kerr-Newman-NUT black holes with a cosmological constant \cite{21}, and multi-black holes \cite{18}. Shadows  for black holes in modified gravity \cite{Moffat}, and the five-dimensional rotating Myers-Perry black hole \cite{Papnoi} have also been investigated. The effect of scattering on the visibility of the shadow was numerically demonstrated by Falcke, Melia and Agol \cite{2}. The visual appearance of an accretion disk was studied with the help of various ray-tracing programs by several authors \cite{3,4,5,6}. The overview of observations and the simulations of phenomena for the black hole in the centre of our galaxy was studied in \cite{7}. 

Most of these investigations of black hole shadows are for idealized cases, dealing with exact solutions in the absence of matter.  It is clearly desirable to seek alternatives to this case, particularly those for which external sources of matter influence the black hole.  Indeed, to understand the properties of black holes as astrophysical objects predicted by the theory of general relativity, it is essential to study their interaction with external matter sources. 
 A particularly relevant example from astrophysics is that of a black hole in a binary system, where  tidal effects from its partner distort the black hole. Although the best way of studying such dynamical systems is numerical analysis, obtaining the most general dynamical exact black hole solution interacting with a general matter configuration is
out of reach. However considerable insight can be gained from studying exact solutions describing a black hole distorted by external matter. Distorted black holes have been of considerable interest for quite some time \cite{Geroch, Israel,Chandrasekhar,Tomimatsu,Breton:1997,Peters,Xanthopoulos,Fairhurst,ASF,ASP}. Such solutions approximate those of dynamical black holes that relax on a time scale much shorter than that of the external matter. 

Here we investigate  how the shadow of a distorted black hole gets deformed. The space-time we consider not asymptotically flat because the sources of distortion are exterior to the black hole region we study: the  observer of the shadow  is not located at asymptotic infinity, but rather at a finite distance from the black hole,  in a vacuum region that is interior to the external sources. We are interested in situations where the distorting potential does not dominate over the potential of the black hole. For this reason we restrict ourselves to the study of small distortions, and for simplicity we consider only the simplest such types. We define the concept of a `local shadow', a shadow seen by an observer not located at asymptotic infinity\footnote{The shadow for an observer not at infinity has been considered in \cite{21}. However, our definition of local shadow is different.}. 
One of our key results is that of relating the deformations of the local shadow of the black hole to its distortion parameters such as the quadruple and octupole multiple moment of the distortion fields. We also find eyebrow structures for a single distorted black hole, similar to that which occurs for two merging black holes \cite{merge}.

In Section 2, we first review the solution representing a distorted axisymmetric Schwarz\-schild black hole. In Section 3, we construct the null geodesic equations for this space-time. In Section 4, we construct a map which would enable us to define the local shadow. In Section 5, we present and discuss our results, and finish with Section 6 where, we sum up.  In what follows employ units where $G=c=1$.

\section{Metric}
We consider a Schwarzschild black hole in an external gravitational field. The solution is static and axisymmetric and its metric can be presented in the following form \cite{Geroch},
\ba\n{St1}
d\mathcal{S}^{2}&=&-\left(1-\frac{2M}{\rho}\right)e^{2U}d\mathcal{T}^{2}\nonumber\\
&+&\left(1-\frac{2M}{\rho}\right)^{-1}e^{-2U+2V} d\rho^{2}\nonumber\\
&+&e^{-2U}\rho^{2}(e^{2V}d\theta^{2}+\sin^{2}\theta d\phi^{2})\,,
\ea
Here, $U=U(\rho,\theta)$ and $V=V(\rho,\theta)$ are functions of $\rho$ and $\theta$.  For $U=V=0$, this metric represents a Schwarzschild black hole. The null rays in this space-time have four conserved quantities, energy $\mathcal{E}$, angular momentum $\tilde{L}$, the azimuthal angular momentum $\tilde{L}_z$, and $u_\mu u^\mu=0$, where $u^{\mu}=dx^{\mu}/d\tau$ is the photon 4-velocity, with $\tau$  an affine parameter along a null geodesic. 
It is convenient to write the metric \eq{St1} in dimensionless form,
\ba
dS^{2}&=&-\left(1-\frac{1}{r}\right)e^{2U}dT^{2}+\left(1-\frac{1}{r}\right)^{-1}e^{-2U+2V} d\rho^{2}\nonumber\\
&+&e^{-2U}r^{2}(e^{2V}d\theta^{2}+\sin^{2}\theta d\phi^{2})\,
\ea
where
\ba
d\mathcal{S}^{2}=r_g^2dS^{2},~~~ T=\frac{\mathcal{T}}{r_g}, ~~~\rho={r_g}r, \n{con1}
\ea 
with $r_g=2M$. The explicit form of $U$ and $V$ as a functions of $r$ and $\theta$ are
\ba
U&=&\sum_{n\ge 0} c_nR^nP_n\nonumber\\
V&=&\sum_{n\ge 1}c_n\sum_{l\ge 0}^{n-1}\left[\cos{\theta}-(2r-1)\right. \nonumber\\
&&\left. -(-1)^{n-l}((2r-1)+\cos{\theta})\right]R^{l}P_l \nonumber\\
&+& \sum_{n,k\ge 1}\frac{nkc_nc_k}{n+k}R^{n+k}(P_nP_k-P_{n-1}P_{k-1})
\ea where, $P_n$ are the Legendre polynomials of the first kind\footnote{Legendre polynomials are solutions to Legendre's differential equation,
\be
\frac{d}{dx}\left[(1-x^2)\frac{d}{dx}P_n(x)\right]+n(n+1)P_n(x)=0 \nonumber\\ \, .
\ee They can be written by using Rodrigues' formula,
\be
P_n(x)=\frac{1}{2^nn!}\frac{d^n}{dx^n}\left[(x^2-1)^n\right] \nonumber\\ \, .
\ee},
\ba
P_n&=&P_n((2r-1)\frac{\cos{\theta}}{R})\nonumber\\
R&=&\sqrt{(2r-1)^2-\sin^2{\theta}}.
\ea 
Moreover, the metric can be written in the form 
\ba
dS^2& =& e^{-2u_{0}}\left[-(1-\frac{1}{r})e^{2\mathcal{U}}dt^2 +{(1-\frac{1}{r})}^{-1}{e^{-2\mathcal{U}+2V}}dr^2\right.\nonumber\\
&&+\left.e^{-2\mathcal{U}}r^{2}(e^{2V}d\theta^2+\sin^2{\theta}d\phi^2) \right],\n{ST3}
\ea
where
\be
\mathcal{U}={U}-u_0,~~~ u_0=\sum_{n\ge 0}c_{2n},~~~ t=Te^{2u_0}\n{STcomformal}
\ee In order to avoid the conical singularities in the space-time we must have
\be
\sum_{n\ge 0}c_{2n+1}=0. \n{nocon}
\ee
In what follows, we construct the null geodesic equations of the following metric
\ba
ds^2& =& -(1-\frac{1}{r})e^{2\mathcal{U}}dt^2 +{(1-\frac{1}{r})}^{-1}{e^{-2\mathcal{U}+2V}}dr^2\nonumber\\
&&+e^{-2\mathcal{U}}r^{2}(e^{2V}d\theta^2+\sin^2{\theta}d\phi^2) ,\n{ST4}
\ea
Note, that metric \eq{ST4} is related to \eq{St1} by conformal transformations  and redefinitions of the time coordinate. The null geodesic equations are conformally invariant.

\section{Null geodesics}

To construct the shape of the shadow of the space-time \eq{ST4}, we need to study the null geodesics for this space-time. 
As with Schwarzschild space-time, the space-time (\ref{ST4}) possesses two commuting Killing vectors
\be\n{Kil}
\xi_{(t)}^{\mu}=\delta^{\mu}_{t},~~~~
\xi_{(\phi)}^{\mu}=\delta^{\mu}_{\phi},
\ee
corresponding to the symmetries of time translation and rotation around the $z$ axis. The corresponding quantities conserved along geodesics of the space-time are 
\ba
E &\equiv&-u_{\mu}\xi^{\mu}_{(t)}=\left(1-\frac{1}{r}\right)e^{2\mathcal{U}}\,\dot{t}\,,\nonumber\\
L_{z} &\equiv&u_{\mu}\xi^{\mu}_{(\phi)}=e^{-2\mathcal{U}}r^{2}\sin^{2}\theta\,\dot{\phi} \n{Con}
\ea
respectively, the photon energy $E$ and azimuthal angular momentum $L_{z}$.
Here $u^{\mu}=dx^{\mu}/d\tau$ is the photon 4-velocity,  normalized as $u^{\mu}u_{\mu}=0$, with $\tau$  an affine parameter along a null geodesic. The geodesic equation is 
\be\n{Geo}
\frac{D^{2}x^{\mu}}{d\tau^{2}}=\frac{d^2 x^\mu}{d\tau^2}+\Gamma_{\alpha\nu}^\mu \frac{d x^\alpha}{d\tau}
\frac{d x^\nu}{d\tau}=0\,,
\ee
$\Gamma_{\alpha\nu}^\mu$ are the Christoffel symbols defined in terms of the metric tensor. 
From \eq{Con}, we see that $d\phi/d\tau$ has  monotonic behaviour, never changing  sign along a particular trajectory.

Therefore, we introduce 
\ba
\dot{r}&=&\frac{dr}{d\tau}=\frac{dr}{d\phi}\frac{d\phi}{d\tau}=r'\dot{\phi}\, , \nonumber\\
\dot{\theta}&=&\frac{d\theta}{d\tau}=\frac{d\theta}{d\phi}\frac{d\phi}{d\tau}=\theta'\dot{\phi} \, . \n{devchange}
\ea
Here, the overdot denotes the derivative with respect to the affine parameter, and prime defines the derivate with respect to $\phi$. 
The geodesic equations take the following form
\ba
r''&=&-\frac{r'^{2}}{2}\left[\frac{h_{,r}}{h}+2\frac{f_{,r}}{f}+\frac{3-4r}{r(r-1)}\right]\nonumber\\
&+&\frac{\theta'^{2}}{2}(r-1)\left[r \frac{h_{,r}}{h}+2\right]\nonumber\\
&+& \theta' r' \left[-\frac{f_{,\theta}}{f}-\frac{h_{,\theta}}{h}+2\cot\theta\right]\nonumber\\
&-&\frac{r^{3}\sin^{4}\theta}{2l_{z}^{2}hf^{3}(r-1)}\left[1+r(r-1)\frac{f_{,r}}{f}\right]\nonumber\\
&-&\frac{1}{hf}\sin^{2}\theta (r-1)\left[\frac{rf_{,r}}{2f}-1\right].\\
\theta''&=&\frac{r'^{2}h_{,\theta}}{2r(r-1)h}
+\frac{\theta'^{2}}{2}\left[- \frac{h_{,\theta}}{h}-2\frac{f_{,\theta}}{f}+4\cot\theta\right]\nonumber\\
&-&\theta' r' \left[\frac{h_{,r}}{h}+\frac{f_{,r}}{f}\right]\nonumber\\
&-&\frac{r^{3}\sin^{4}\theta f_{,\theta}}{2l_{z}^{2}hf^{4}(r-1)} - \frac{\sin^{2}\theta}{2hf}\left[\frac{f_{,\theta}}{2f}-2\cot\theta\right]
\ea
where, $f=e^{2\mathcal{U}}$, $h=e^{-2\mathcal{U}+2V}$, $r''=d^{2}r/d\phi^{2}$, $\theta''=d^{2}\theta/d\phi^{2}$, and we have defined $l_{z}={L_{z}/E}$.
We also have the following constraint
\ba
&&u_{\mu}u^{\mu}=0\, , \nonumber\\
&&-\frac{r^4\sin^4\theta}{f^2l_z^2}+r^2\sin^2\theta F+fhr'^2+r^2Ffh\theta'^2=0 \n{umuum} \nonumber\\
\ea
where  $F=1-1/r$.

In order to study the effect of distortions on the black hole, we consider the simplest cases. The simplest type of distortion is due to a monopole $c_0\neq 0$ and $c_{n}=0$ for $n>0$. The next less trivial distortion is due to a dipole $c_0\neq0$ and $c_{1}\neq 0$. In this case from condition (\ref{nocon}), we have $c_{1}=-c_{3}$, and we consider $c_n=0$ for any other $n$. The other case is the quadrupole moment where $c_0\neq 0$ and  $c_{2}\neq 0$ and $c_{n}=0$ for any other case. We shall consider the dipole and quadrupole cases separately. Recall that since we made the conformal transformation (\ref{ST3}-\ref{STcomformal}) the quantity $c_0$ does not contribute to our result. 

Here, we have assumed that in analogy to the Newtonian multiple moments we have $c_0>c_1>c_2>...>c_n$.  Note that if we calculate the multiple moments for a static axisymmetric Newtonian gravitational potential 
of two point-like masses $M_1$ and $M_2$ and a ring located on the equator of the axis joining the two masses, we find that $c_0>c_1>c_2>...>c_n$, and the ring contributes only to the even multiple moments. In this sense a quadrupole multiple moment $c_2$ is analogous to that of a ring around the black hole.
For the quadrupole distortion we have
\ba
V& =&-2c_2\sin^2{\theta}\left[2r-1+18c_2r^4\cos^2{\theta}-36c_2r^3\cos^2{\theta} \right. \nonumber\\
&+& 22c_2r^2\cos^2{\theta}-4c_2r\cos^2{\theta}-2c_2r^4 \nonumber\\
&+& \left. 4c_2r^3-2c_2r^2\right], \\
\mathcal{U}&=&-c_2+\frac{c_2}{2}(2r-1)^2(3\cos^2\theta-1)+\frac{c_2}{2}\sin^2\theta.
\ea
For the dipole distortions (odd multiple moments) we have
\ba
V& =& A\cos{\theta}+A_0\cos^2{\theta}+A_1\cos^4{\theta}+A_2
\nonumber\\
\mathcal{U}&=&-c_1\cos{\theta}\left[r\left[\cos^2{\theta}(20r^2-30r+12)\right.\right. \nonumber\\
&-&\left.(12r^2-18r+8)\left. \right]+\sin^2{\theta}\right]
\ea 
where the functions  $A$, $A_0$, $A_1$, $A_2$ of $(r,\theta)$ are given  by
\ba
A& =& 2c_1\sin^2{\theta}\left[6r(r-1)+1\right] \, ,\nonumber\\
A_0&=& -2{c_1}^2r\sin^2{\theta}\left[-168r^5+504r^4-570r^3\right.  \nonumber\\
&+&\left. 300r^2-72r+6\right] \, , \nonumber\\
A_1&=& -2{c_1}^2r\sin^2{\theta}\left[300r^5-900r^4+1008r^3\right.  \nonumber\\
&-&\left. 516r^2+117r-9\right] \, , \nonumber\\
A_2&=& -2{c_1}^2r\sin^2{\theta}\left[12r^5-36r^4+42r^3 \right. \nonumber\\
&-& \left. 24r^2+7r-1\right] \, .
\ea

\section{Map}

\begin{figure}
\begin{center}
\includegraphics[width=6cm]{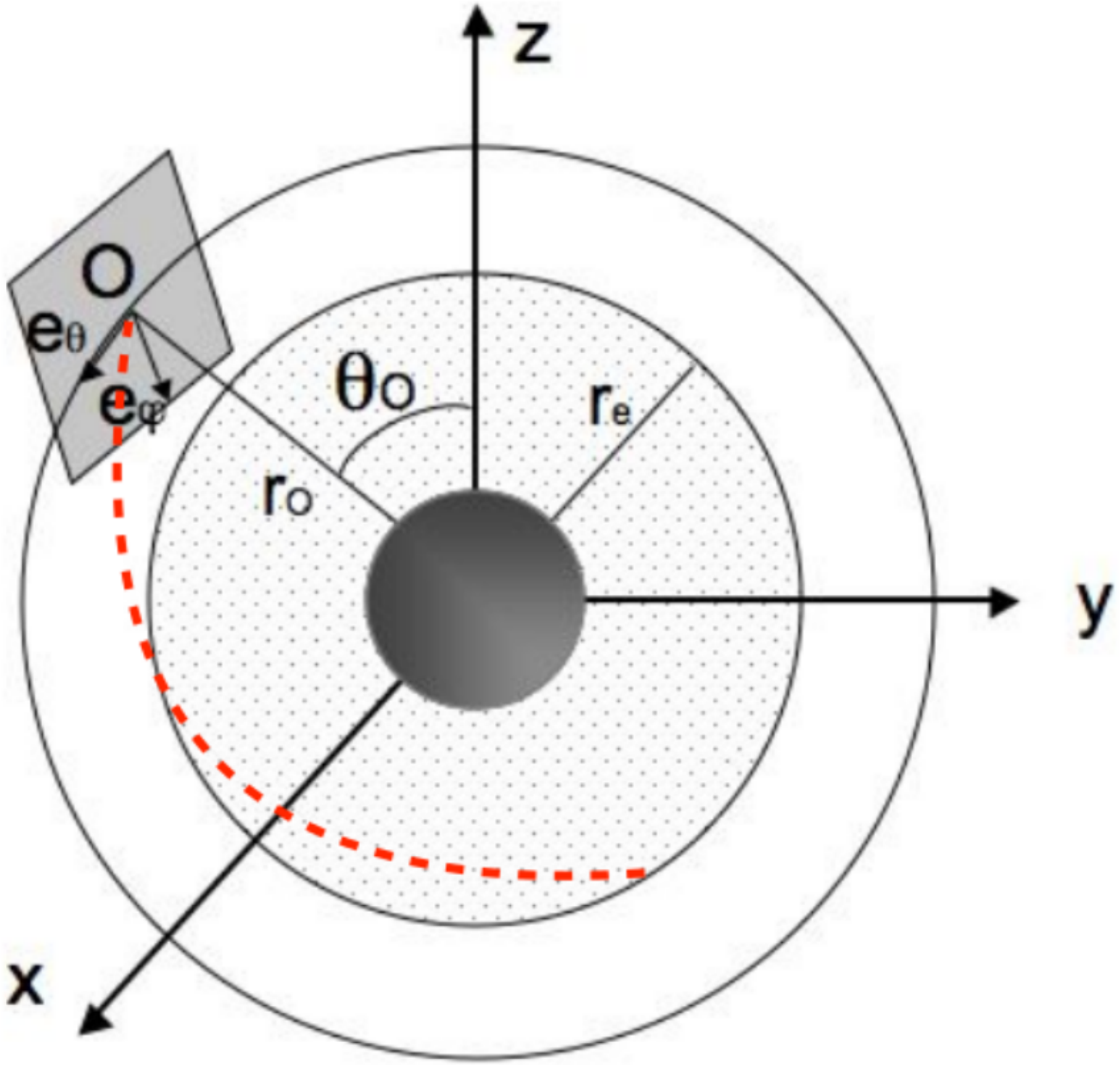}
\caption{ The photon emitted from a point on the sphere of radius $r_{e}$ reaches an observer located at $(r_{o},\theta_{o})$. ${\bf e}_{\theta}$ and ${\bf e}_{\phi}$ are unit vectors of the observer's orthonormal tetrad directed along the coordinate lines of $\theta$ and $\phi$. The red dashed line represents an example of a null trajectory that reaches the observer.}\label{Fig1}
\end{center}
\end{figure}

Our goal is to see how an observer located at $(r_{o},\theta_{o})$ sees the shadow of a distorted black hole. Consider rays emitted from any point on a sphere of radius $r_{e}$ around the black hole that may reach the observer located at $(r_{o},\theta_{o})$ or else be absorbed by the black hole as shown in  figure \ref{Fig1}.   This observer is located in the interior region, which means that the sources responsible for the distortion of the black hole are all located at $r  > r_{o} \geq r_{e}$.

To analyze the problem  we consider   rays  emitted at $(r_{o},\theta_{o})$ and trace them backward to their source. In order to characterize properties of  rays emitted at  $(r_{o},\theta_{o})$, which are directly connected with observation, we proceed as follows. First we choose the following orthonormal tetrad 
\ba
&&e^{\mu}_{0}=\frac{1}{\sqrt{(1-\frac{1}{r})f}}\delta^{\mu}_{t} \Big|_{o},~~~
e^{\mu}_{1}=\sqrt{\frac{1}{h}(1-\frac{1}{r})}\delta^{\mu}_{r}\Big|_{o} \, , \nonumber\\
&&e^{\mu}_{2}=\frac{1}{r\sqrt{h}}\delta^{\mu}_{\theta}\Big|_{o}, ~~~
e^{\mu}_{3}=\frac{\sqrt{f}}{r\sin\theta}\delta^{\mu}_{\phi}\Big|_{o}\, . \n{ortho} 
\ea
at the point of the observation,
where $...\Big|_{o}$ stands for the limit where $r=r_{o}$ and $\theta=\theta_{o}$.
These vectors are orthonormal. In what follows, we use $(t,r,\theta,\phi)$ coordinates and the indices are $(0,1,2,3)$, respectively. The vector ${\bf e}_{1}$ is in radial direction to the direction of the black hole. The tangent vector to a null ray is
\be
u^{\mu}=\frac{dx^{\mu}}{d\tau}.\n{tangent}
\ee
We project this vector into the observer's orthonormal frame.  
Thus, the tangent vector at the observation point $(r_{o},\theta_{o})$ can be written as
\be
u^{\mu}=\beta(-e^{\mu}_{0}+\xi^{1} e^{\mu}_{1}+\xi^{2} e^{\mu}_{2}+\xi^{3} e^{\mu}_{3}),\n{tangent2}
\ee
where $\xi^{1}$, $\xi^{2}$, and $\xi^{3}$ are displacement angles.  Replacing \eq{ortho} in  \eq{tangent2} and comparing to \eq{tangent} and using \eq{Con}, we find the value of the scalar coefficient $\beta$
\ba
&&\beta=-\frac{E}{\sqrt{(1-\frac{1}{r})f}}. \n{beta1}
\ea

Consider a photon that gets emitted from a point located on a sphere of radius ${r}_{e}$. Following the motion of this null ray forward, it will either get captured by the black hole or reach the eye of the observer located at $(r_{o},\theta_{o})$ depending on the shape of the black hole.  To determine the shadow of the black hole, we instead trace  the trajectory of these photons  backward (see figure \ref{Fig1}). The quantities $\xi^{2}$, $\xi^{3}$ give the angle of the photon as it reaches the observer located at point $(r_o,\theta_o)$. Suppose  the observer is looking in the direction  of the centre of the black hole.  Tracing straight backward null rays 
leaving the observer at  angles $\xi^{2}$, $\xi^{3}$ as  if the space were flat, it appears to him/her that the photon has reached his/her eye from the point 
$\bar{\xi}^{2}=\xi^{2} r_o$ and $\bar{\xi}^{3}=\xi^{3} r_o$ on the plane of the black hole.  If this null ray is absorbed by the black hole, we consider $(\bar{\xi}^{2}, \bar{\xi}^{3})$ to be a black point on the plane of the black hole -- by definition it is  a member of the ``local shadow'' of the black hole, so named because the observer is not located at asymptotic infinity.  Replacing \eq{ortho} in  \eq{tangent2} and comparing to \eq{tangent}, we find the displacement angles $\xi^{2}$, and $\xi^{3}$ and using \eq{beta1}. Moreover, by multiplying by $r_o$ and using \eq{devchange} we find 
\ba
&&\bar{\xi}^{2}=\pm\frac{l_{z}}{\sin^{2}\theta}f^{\frac{3}{2}}\sqrt{h(1-\frac{1}{r})}\theta'\Big|_{o},\n{t1a}\nonumber\\
&&\bar{\xi}^{3}=-\frac{l_{z}}{\sin\theta}f\sqrt{1-\frac{1}{r}}\Big|_{o}.\n{t2a}
\ea 
On the other hand, if this null ray reaches  a radius $r_{e}$, (after propagating in the space-time) it is not a member of the  local shadow of the black hole. 

Before we close this section of the map we perform a ``test''. We want to see if by taking the limit $r\rightarrow\infty$ and setting the distortions to zero $(f=h=1)$ we will get the map  \cite{11},
\ba
\bar{\xi}^{2}&=&-\dot{\theta}\frac{r^2\sqrt{1-\frac{1}{r}}}{E}\Big|_{r\rightarrow\infty}\, . \n{test1}
\ea At infinity and without distortions the system is completely integrable and the equation of motion for the $\dot{\theta}$ is given in the following form,

\be
\dot{\theta}=-\frac{\sqrt{{L_0}^2\sin^2{\theta}-{L_z}^2}}{r^2\sin{\theta}}\, .
\ee Therefore \eq{test1} takes the form,
\ba
\bar{\xi}^{2}&=&\sqrt{{l_0}^2-\frac{{l_z}^2}{\sin^2{\theta}}} \n{t1} \, .
\ea For $\bar{\xi}^{3}$ we have,
\ba
\bar{\xi}^{3}&=&-\frac{l_z}{\sin{\theta}}\sqrt{1-\frac{1}{r}}\Big|_{r\rightarrow\infty}=
-\frac{l_z}{\sin{\theta}} \, . \n{t2}
\ea 

We see that equations (\ref{t1},\ref{t2}) match with the equations for the map of undistorted black holes \cite{11}.
In the Schwarzschild undistorted space-time,  \eq{St1} for $U=V=0$, there exists a second rank Killing tensor $K_{\mu\nu}$ ($K_{(\mu\nu;\alpha)}=0$) which generates another conserved quantity, the squared total angular momentum of a photon
\be
 {\tilde{L}}^{2}\equiv K^{\alpha\beta}u_{\alpha}u_{\beta}\,.\n{L}
\ee
 For the critical impact parameter $\tilde{b}^2_\text{critical}=\tilde{L}^2/\mathcal{E}^2=27M^2$, the light enters a circular knife-edge orbit of radius $\rho=3M$, and may orbit the black hole for part of an orbit or for many turns before it escapes or plunges. These rays barely escape the Schwarzschild black hole and form the rim of the black hole shadow. For the rescaled metric \eq{ST3} with $\mathcal{U}=V=0$. These rays have $L^2/E^2=27/4$ where, $L=\tilde{L}/r_g$ and $\mathcal{E}=r_gE$. For an observer located at infinity this implies that 
\ba
(\bar{\xi}^2)^2+(\bar{\xi}^3)^2=\frac{27}{4}.\n{ShwShadow0}
\ea
Note that, usually the radius of the shadow of a Schwarzschild black hole as seen by an observer at infinity is $3\sqrt{3}M$. However, in our case, since, $\rho=2Mr$, (see \eq{con1}), the radius of the shadow of a Schwarzschild black hole as seen by an observer at infinity is $\sqrt{27}/4=3\sqrt{3}/2$. From equation (\ref{t2a}), when $f=h=1$, for an observer located at $r_o$, we have 
\ba
(\bar{\xi}^2)^2+(\bar{\xi}^3)^2=\frac{{l_{z}}^2(r_o-1)}{\sin^4 \theta r_o} \left(\theta'^2 +\sin^2\theta\right)
=\frac{27(r_o-1)}{4r_o}\nonumber\\
\n{M1}
\ea 
upon setting $L^2/E^2\equiv l^2=27/4$ and using ${l_{z}}^2\theta' = {\sin\theta}\sqrt{l^2 \sin^2\theta-{l_{z}}^2}$ from
 Schwarzschild space-time.

\section{Results and Discussion}

Here we study the effect of distortions on the local shadow of an undistorted black hole.  Imagine for example that there is a ring around the black hole with radius $R$, and the observer is at a radius $r_o<R$. The distortions are characterized in terms of the multiple moments $c_n$. For simplicity we consider the lowest order multiple moments (considering  odd and even  moments  separately). In choosing the appropriate value of multiple moment, we suppose its value to be small enough that the ratio $f=g_{tt\text{d}}/g_{tt}$ of the $tt$-component of the distorted black hole metric to its Schwarzschild counterpart is not very large, i.e.,  $f < 10$. This ratio is illustrated in Figure \ref{Fig_13}. 
 
We have also calculated the $f$ and $h$ functions in figures \ref{Fig_13}-\ref{Fig_12}. We need to consider smaller values of $c_1$ compared to $c_2$. For example, $c_1=1/150$ is a very strong distortion, such that $g_{tt\text{d}}/g_{tt}>10000$. We need to keep in mind that we want to consider a black hole distorted by external sources that are not much stronger  gravitationally than the black hole itself. We consider $-1/150<c_2<1/150$; for this range of values of $c_2$ we have, $g_{tt\text{d}}/g_{tt} < 2.6$. We assume, $-1/800<c_1<1/800$;  we find $g_{tt\text{d}}/g_{tt} < 6$ for this range of values of $c_1=-c_3$.  
 
In the following, we consider only the case where $r_{e}=r_{o}$. In our numerical computations, for every photon trajectory with initial values  $\bar{\xi}^{3}$ and $\bar{\xi}^{2}$, we compute $l_z$, $\theta'$. 
From the constraint equation \eq{umuum} we can derive $r'$ as following
\be
r'=\biggl[ \frac{r^4 \sin^4\theta}{f^3 l_z^2 h}-\frac{r^2 \sin^2\theta}{fh}(1-\frac{1}{r})-r^2\theta'^2(1-\frac{1}{r})\biggl]^{\frac{1}{2}}. \n{rp1}
\ee 
From \eq{rp1}, we see that $\bar{\xi}^{3}$ and $\bar{\xi}^{2}$ cannot take every value. Computing this quantity 
for any choice of 
$(r_{o},\theta_{o})$ yields
 \be
 r'(r_{o},\theta_{o}) =   \frac{r^2 \sin^4\theta}{f^3 l_z^2 h}\Big|_{(r_{o},\theta_{o})}\sqrt{r^2_{o} -
(\bar{\xi}^{3})^2-(\bar{\xi}^{2})^2}
\n{xy}
\ee
and for $\bar{\xi}^{3}$ and $\bar{\xi}^{2}$ outside a circle of radius $r_o$, the expression under the root becomes negative. Thus relation \eq{xy} gives us the range of values of $\bar{\xi}^{3}$ and $\bar{\xi}^{2}$ that are valid. Numerically, we divide the interior region of the circle of \eq{xy} to small pixels of side 0.01; for each of these values we compute the corresponding photon trajectory. We see whether this trajectory ends inside the black hole or escape the black hole, namely reaches  radius $r_{o}$. For each trajectory, we  calculate the maximum deviation from the constraint \eq{umuum}. From this we  find the maximum deviation from the constraint for all the trajectories, which we find to be $4\times 10^{-4}$.
\begin{figure}[htb]
\begin{center}
\ba
 &&\includegraphics[width=6cm]{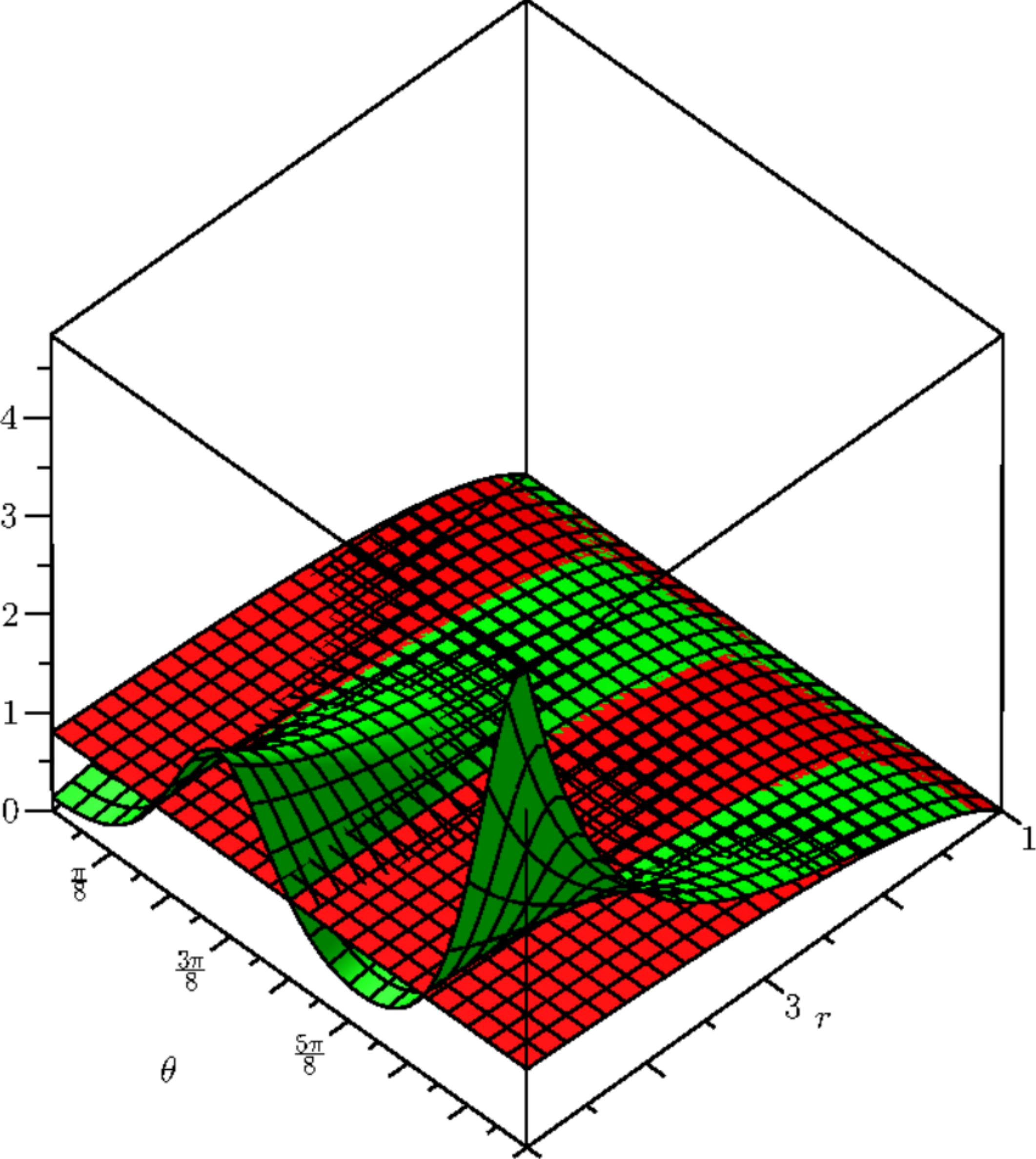}\nonumber\\
 && \includegraphics[width=6cm]{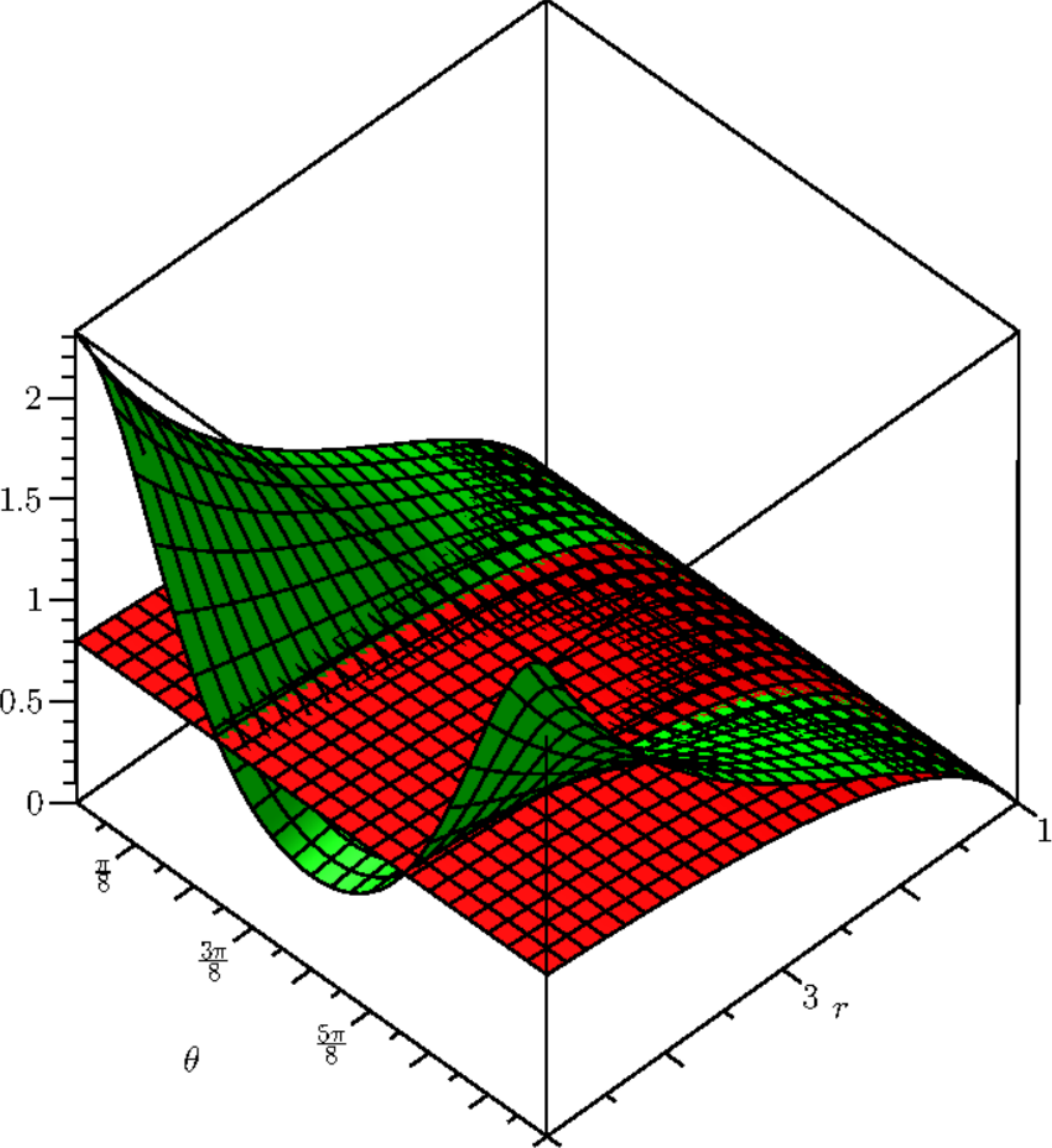}
\ea
  \caption{  Comparison of $g_{ttd}$ of the distorted black hole (black), to  $g_{tt}$ of the Schwarzschild black hole (grey). 
  Top: $c_1=1/800$; 
  bottom for $c_2=1/150$.  
  \label{com}}
  \end{center}
\end{figure}
\begin{figure}[htb]
\begin{center}
\ba
 &&\includegraphics[width=6cm]{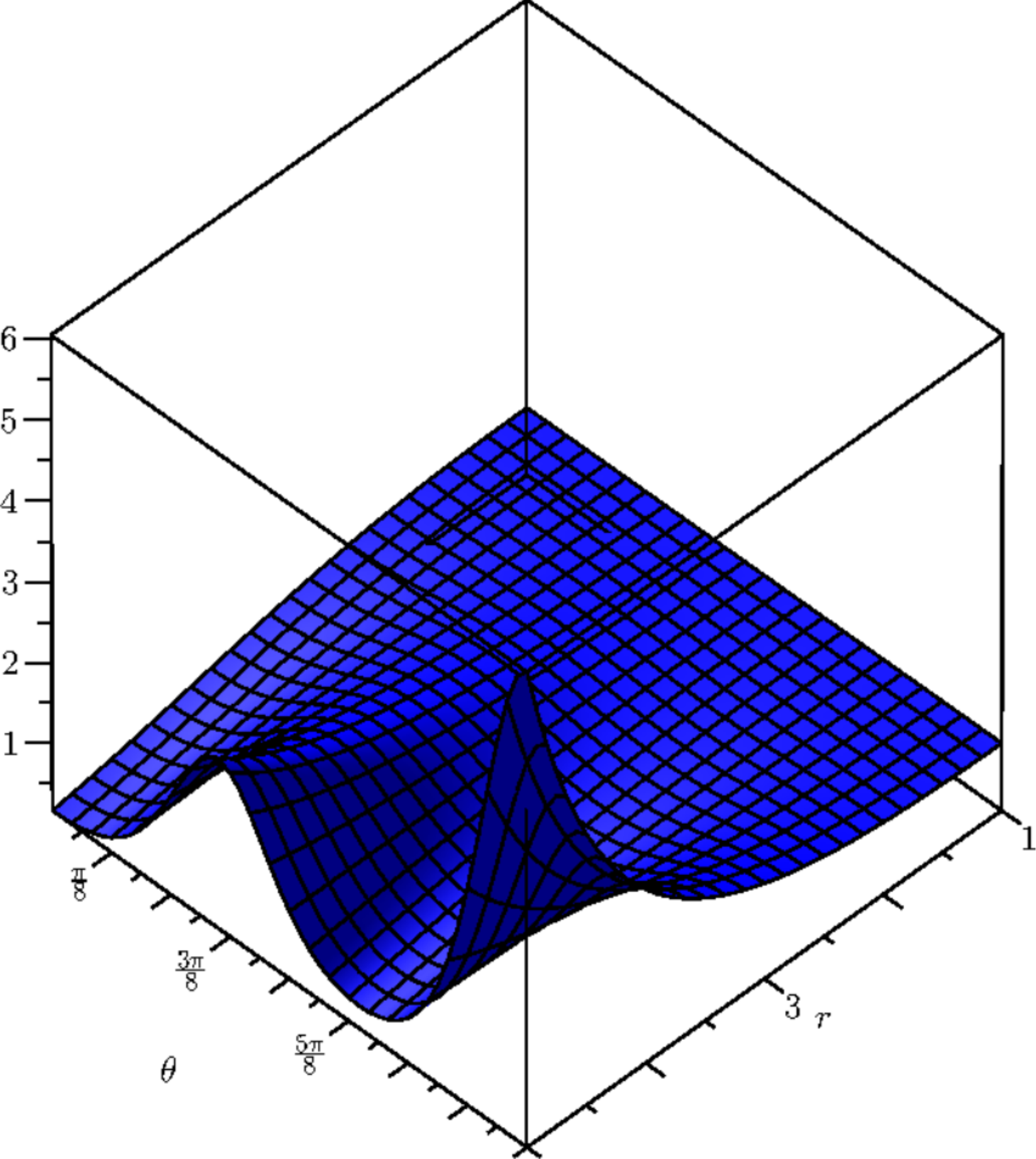}\nonumber\\
 &&\includegraphics[width=6cm]{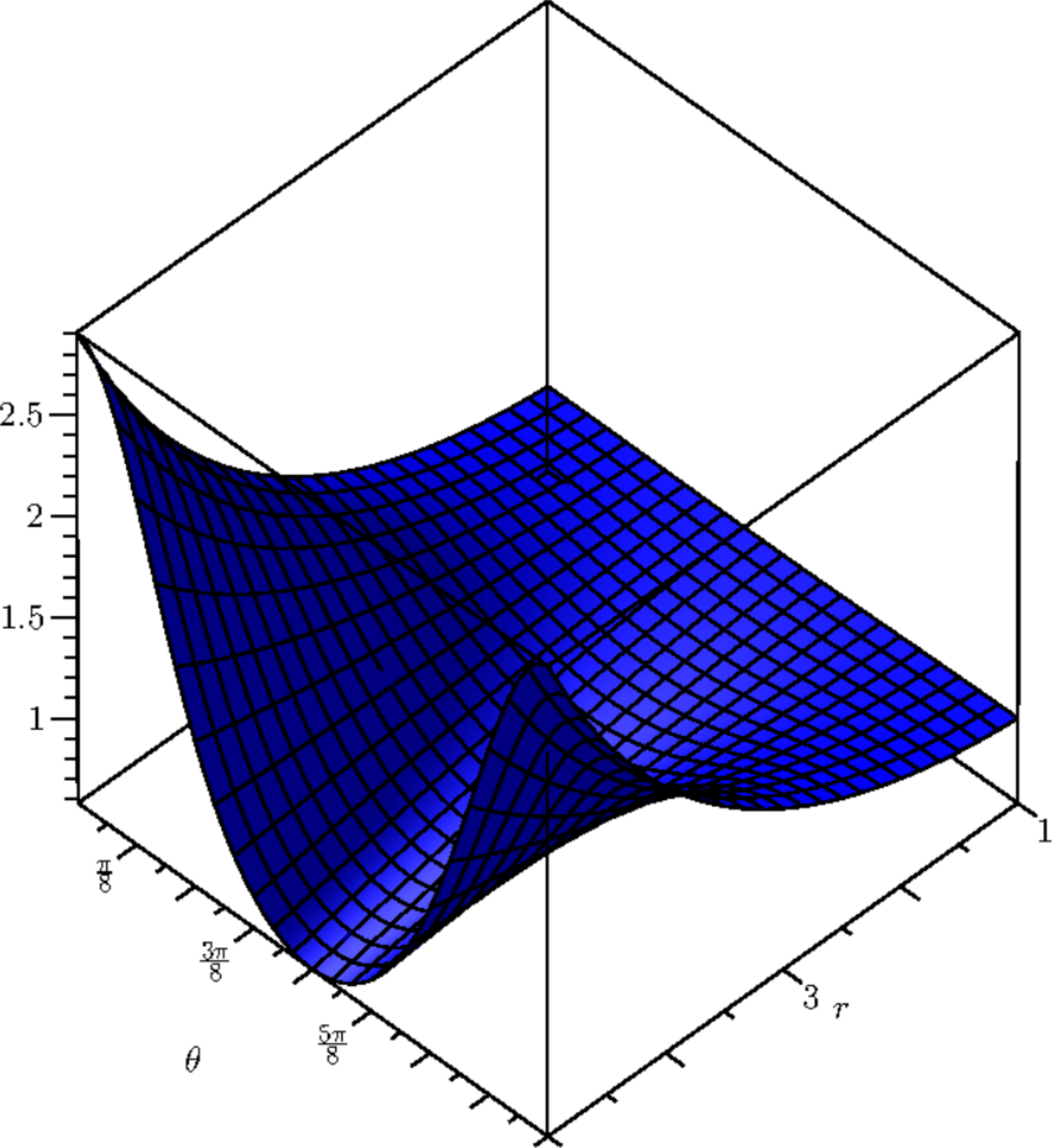}
  \ea
    \caption{ $f$ function for distorted black hole.  Top: $c_1=1/800$; Bottom $c_2=1/150$ }\label{Fig_13}  
  \end{center} 
\end{figure}
\begin{figure}[htb]
\begin{center}
\ba
 &&\includegraphics[width=6cm]{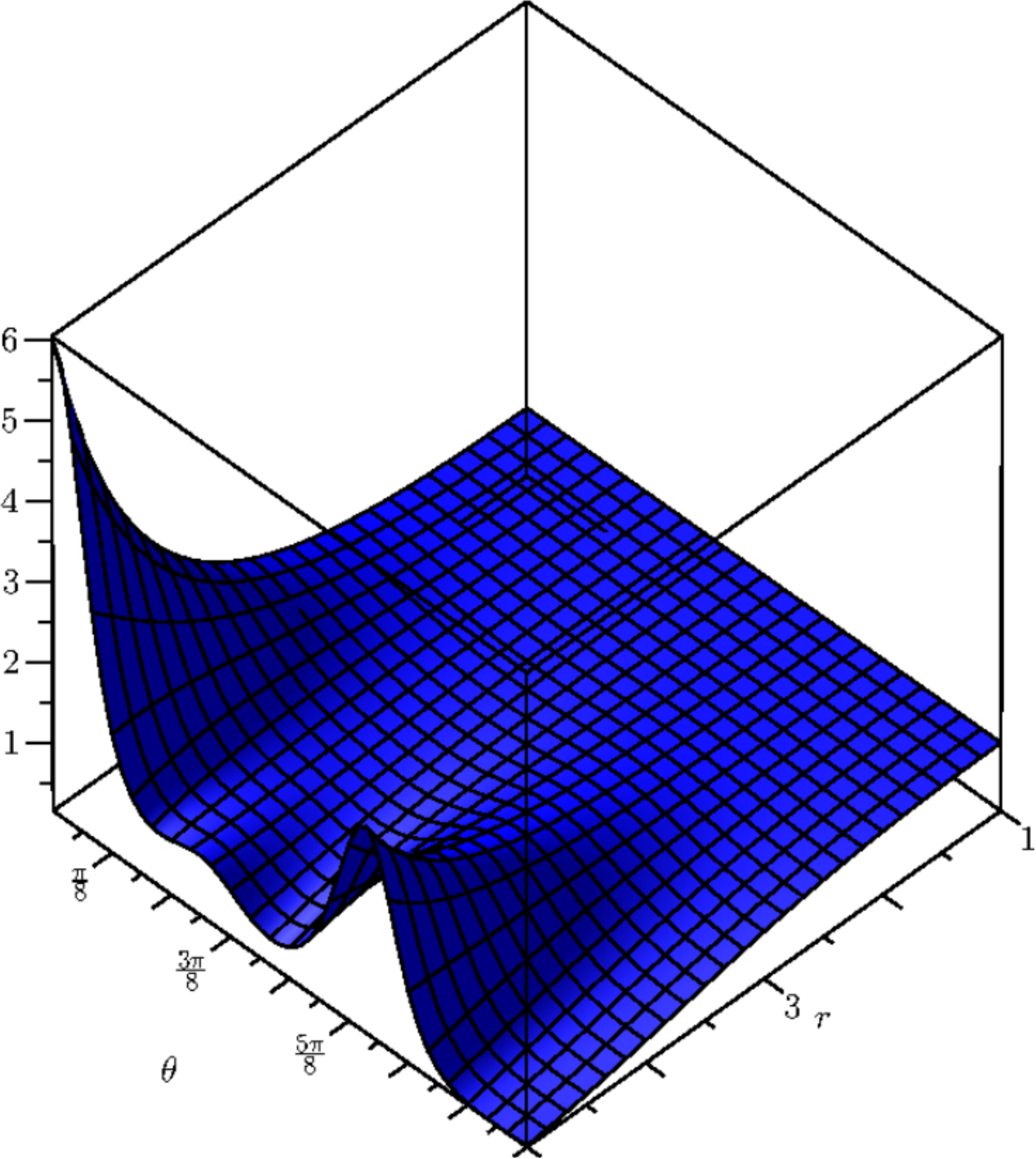}\nonumber\\
&&\includegraphics[width=6cm]{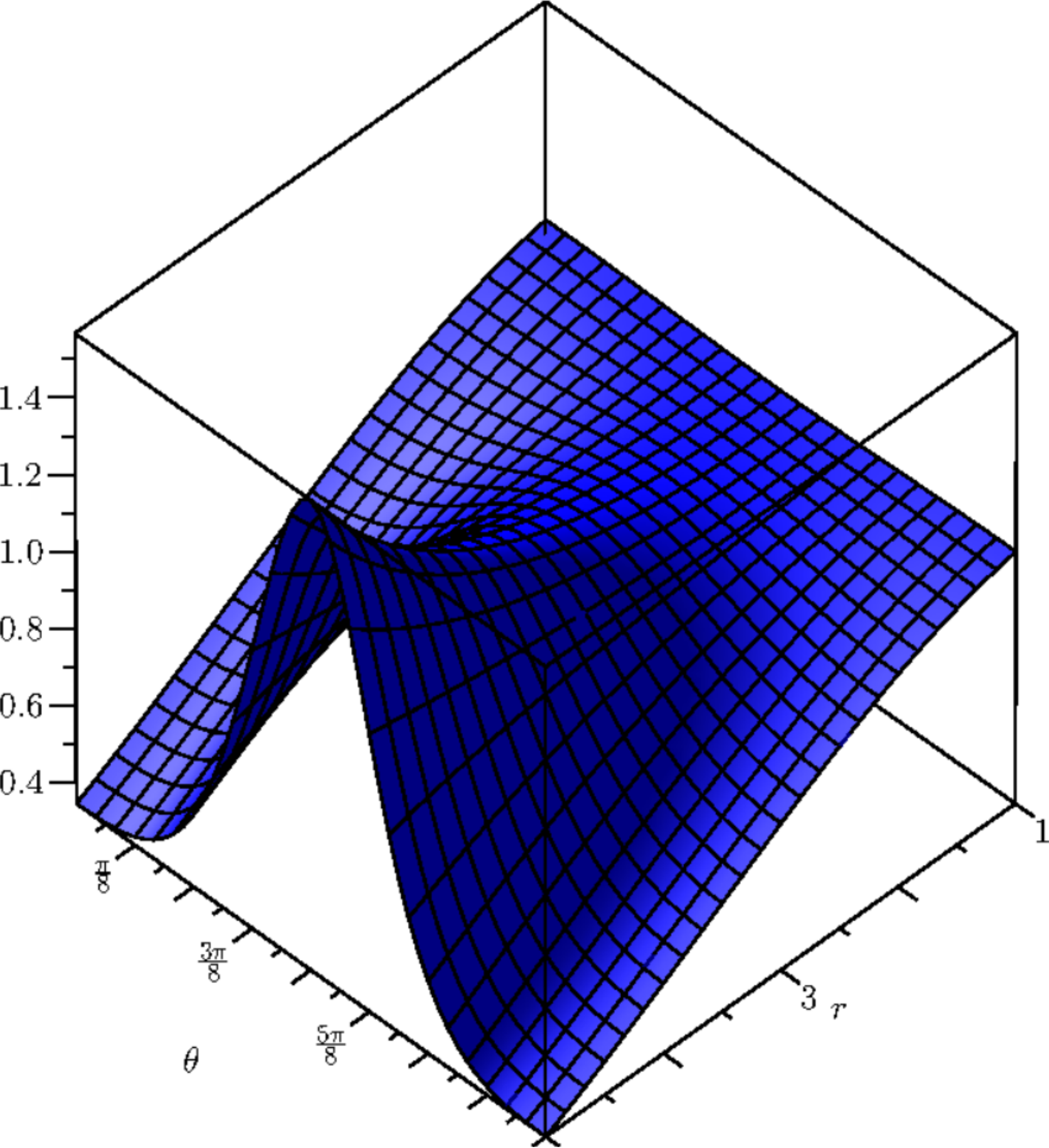}
  \ea
    \caption{ $h$ function for distorted black hole.  Top: $c_1=1/800$; Bottom:  $c_2=1/150$. }
\label{Fig_12}  
\end{center} 
\end{figure}
\begin{figure}[htb]
\begin{center}
\ba
&&\includegraphics[width=6cm]{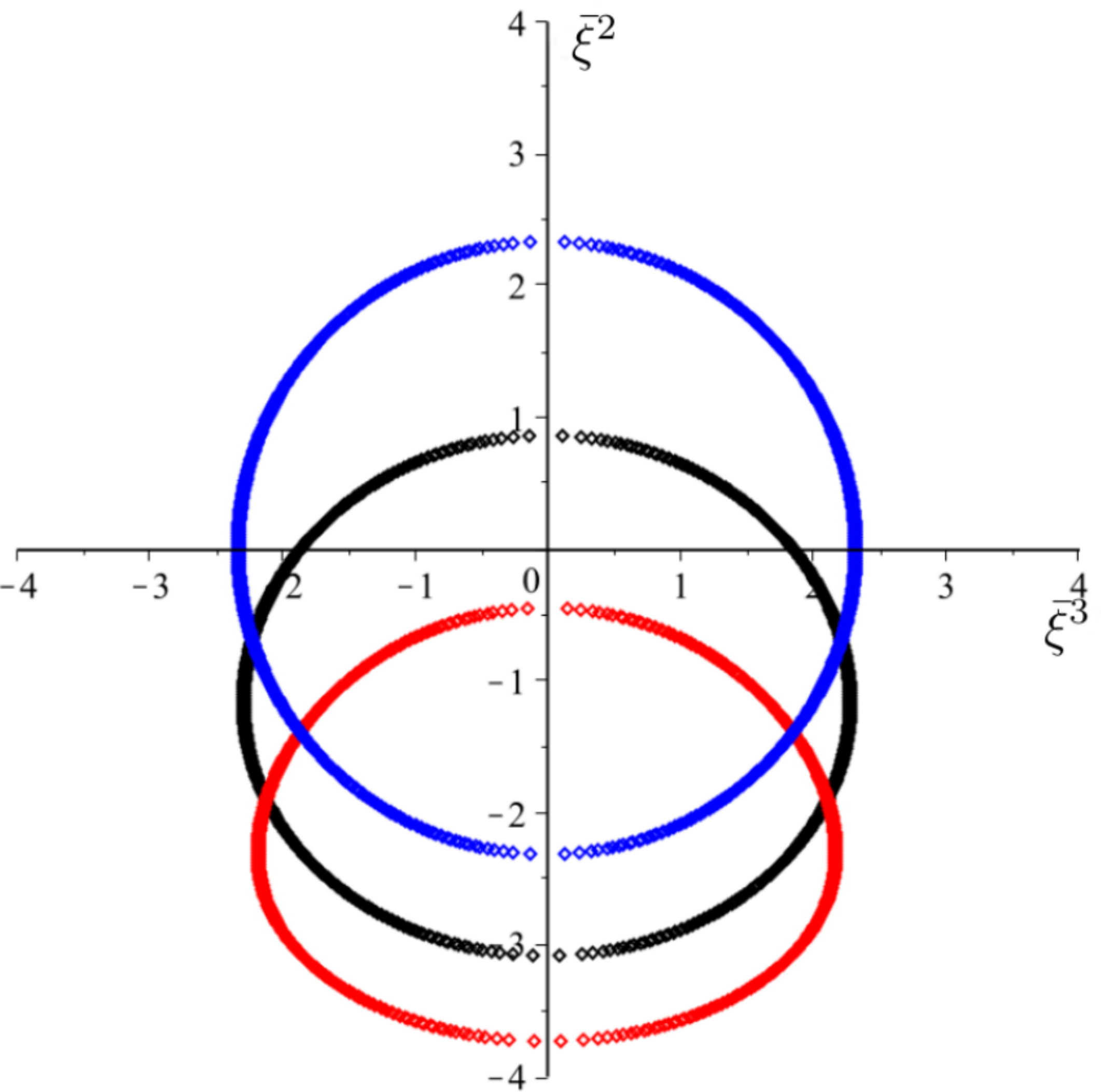}\nonumber\\
&&\includegraphics[width=6cm]{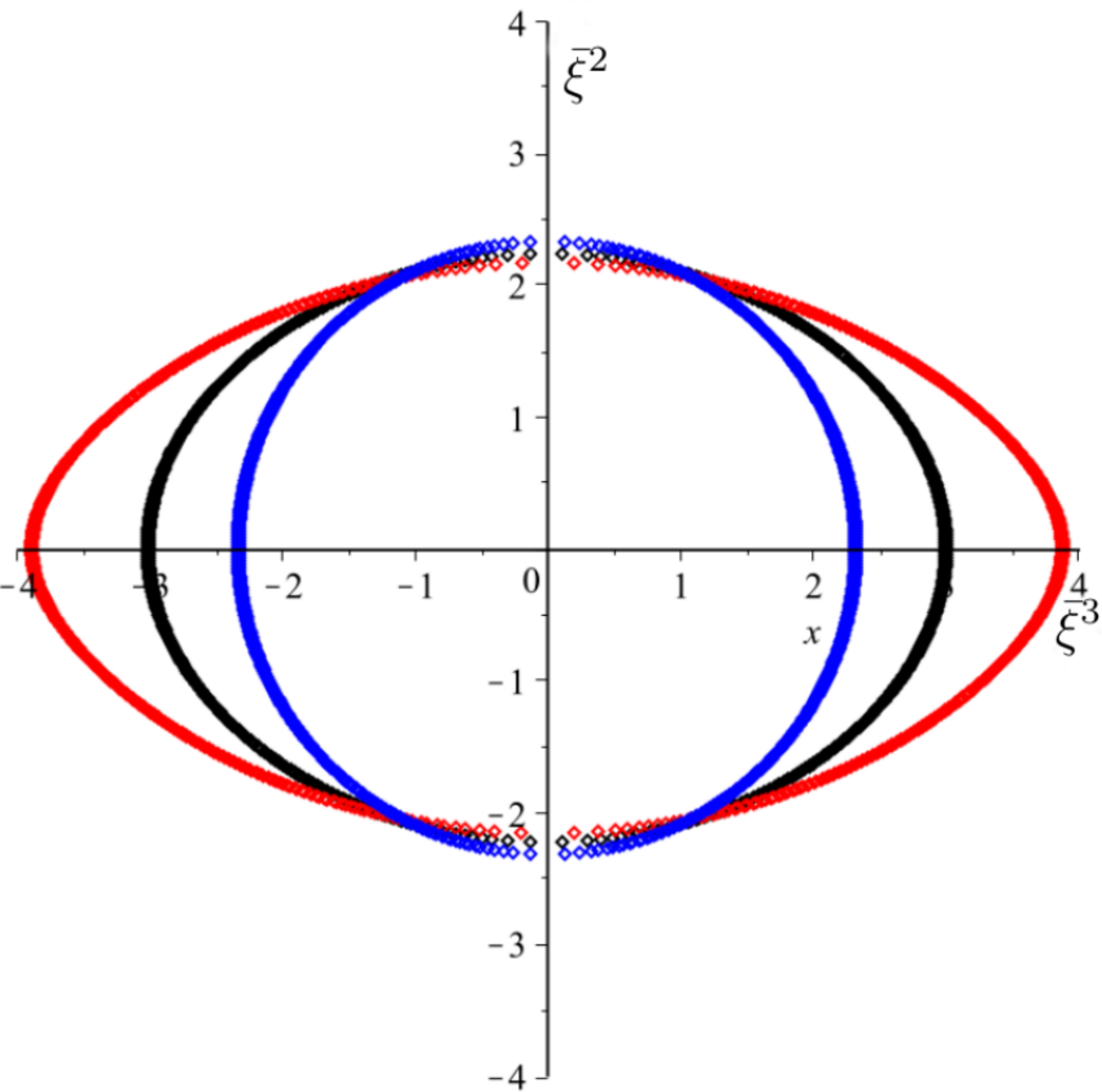}
  \ea
  \caption{ \label{Fig_5}  On the top we can see the shadow of a black hole for an observer at $\theta_o=\pi/4$ and radius $r_o=5$; the values of the multiple moments are $c_2=-\frac{1}{150}$, (with red, (3)) $c_2=-\frac{1}{300}$, (with black, (2)) and $c_2=0$, (which is the undistorted case with blue, (1)). On the bottom we can see the shadow of a black hole for an observer at $\theta_o=\pi/2$ and radius $r_o=5$; the values of the multiple moments are $c_2=-\frac{1}{150}$, (with red, (3)) $c_2=-\frac{1}{300}$, (with black, (2)) and $c_2=0$, (which is the undistorted case with blue, (1)).}
\end{center} 
\end{figure}
\begin{figure}[htb]
\begin{center}
\ba
 &&\includegraphics[width=6cm]{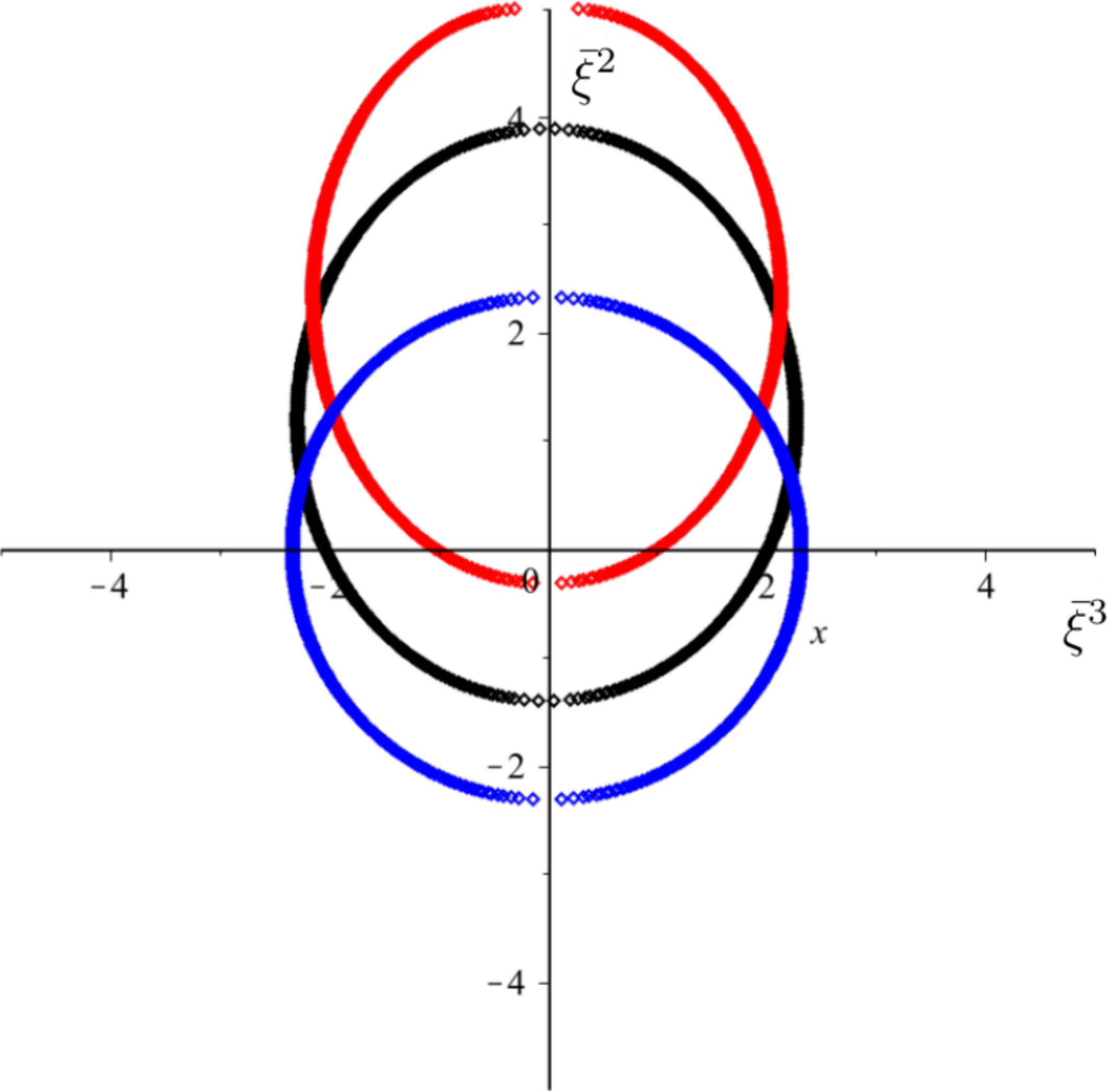}\nonumber\\
 &&\includegraphics[width=6cm]{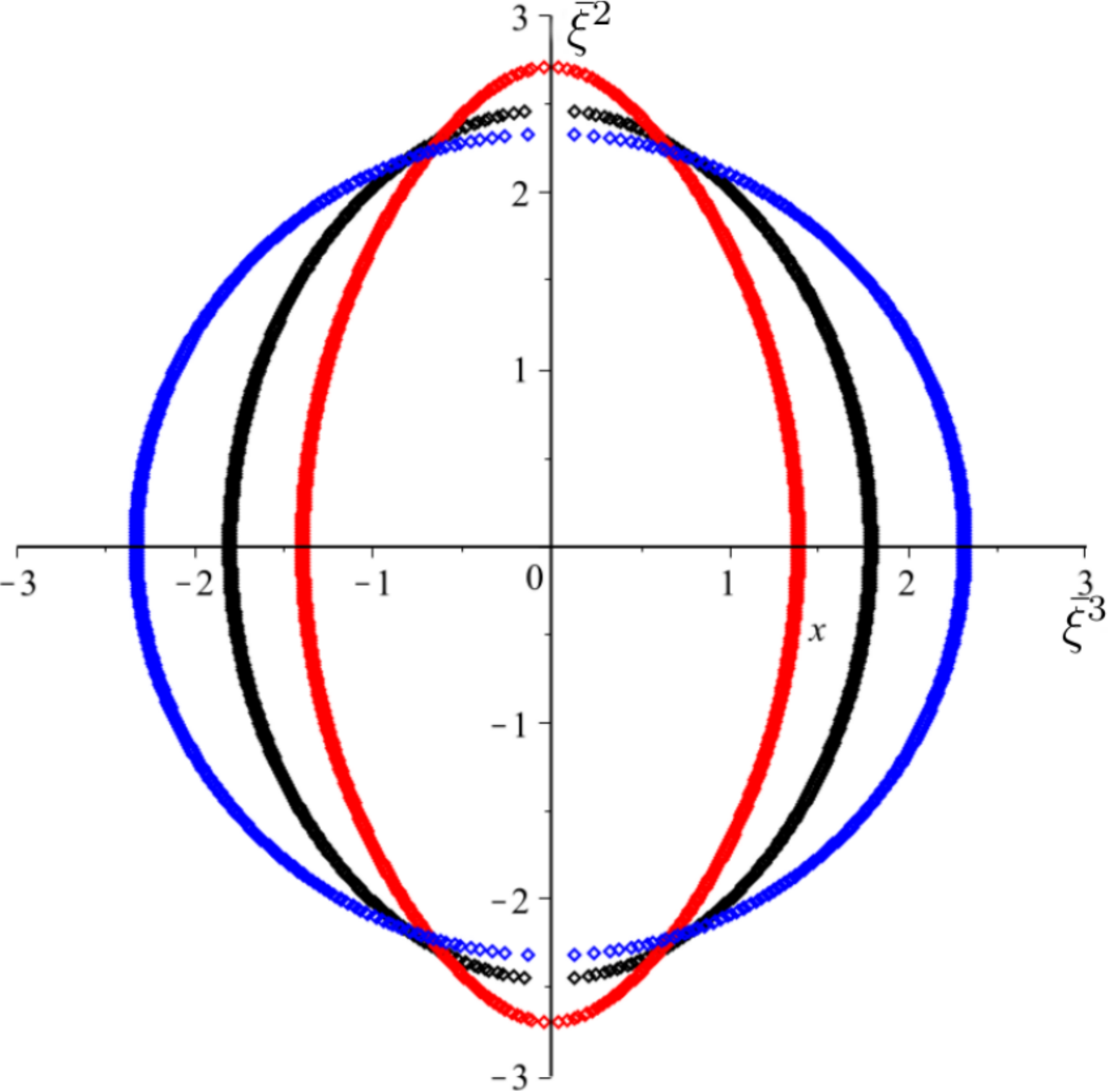}
   \ea
  \caption{ \label{Fig_6} On the top we can see the shadow of a black hole for an observer at $\theta_o=\pi/4$ and radius $r_o=5$; the values of the multiple moments are $c_2=\frac{1}{150}$, (with red, (3)) $c_2=\frac{1}{300}$, (with black, (2)) and $c_2=0$, (which is the undistorted case with blue, (1)). On the bottom we can see the shadow of a black hole for an observer at $\theta_o=\pi/2$ and radius $r_o=5$; the values of the multiple moments are $c_2=\frac{1}{150}$, (with red, (3)) $c_2=\frac{1}{300}$, (with black, (2)) and $c_2=0$, (which is the undistorted case with blue, (1)).}
\end{center}  
\end{figure}
\begin{figure}[htb]
\begin{center}
\ba
 &&\includegraphics[width=6cm]{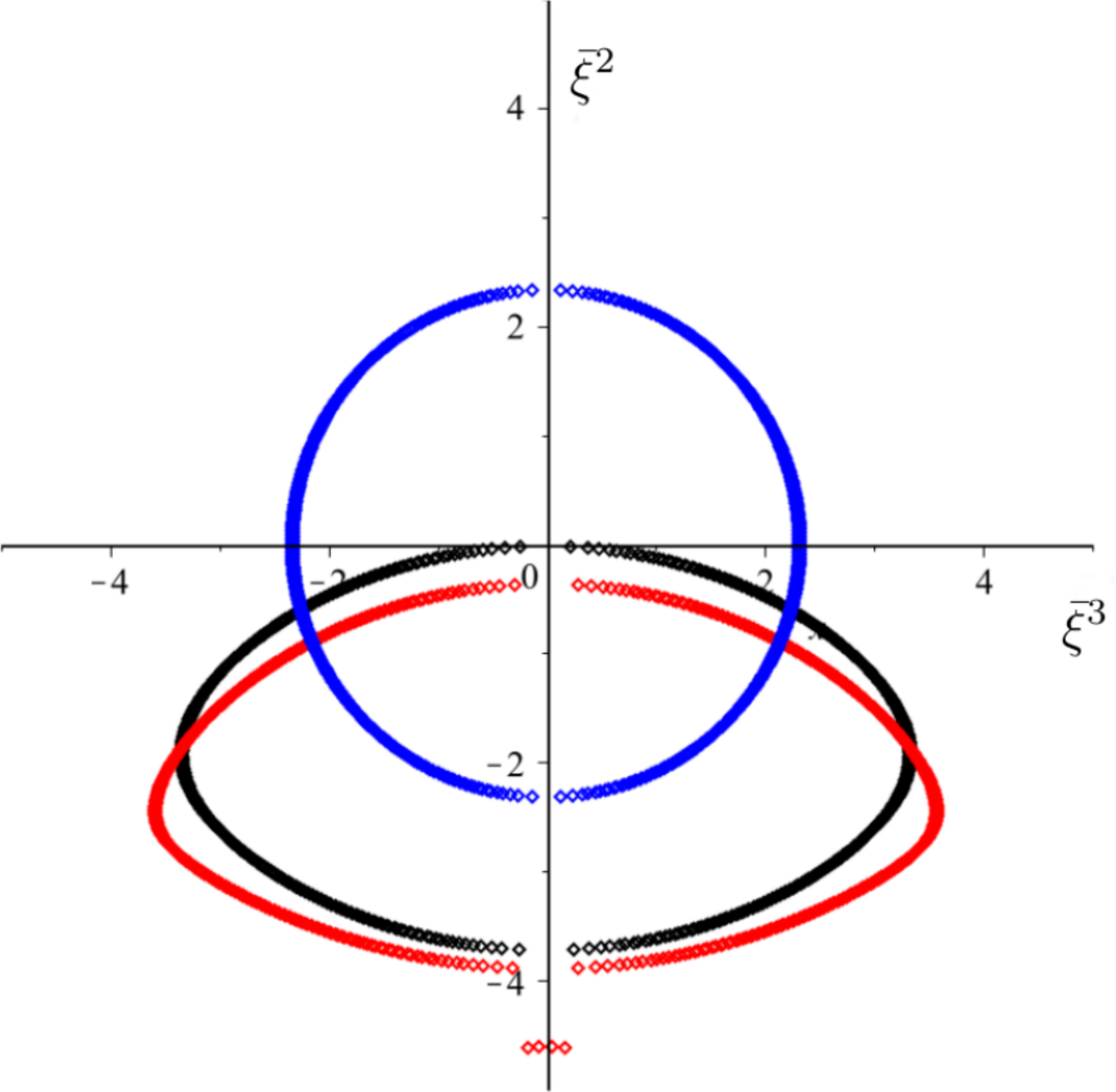}\nonumber\\
 &&\includegraphics[width=6cm]{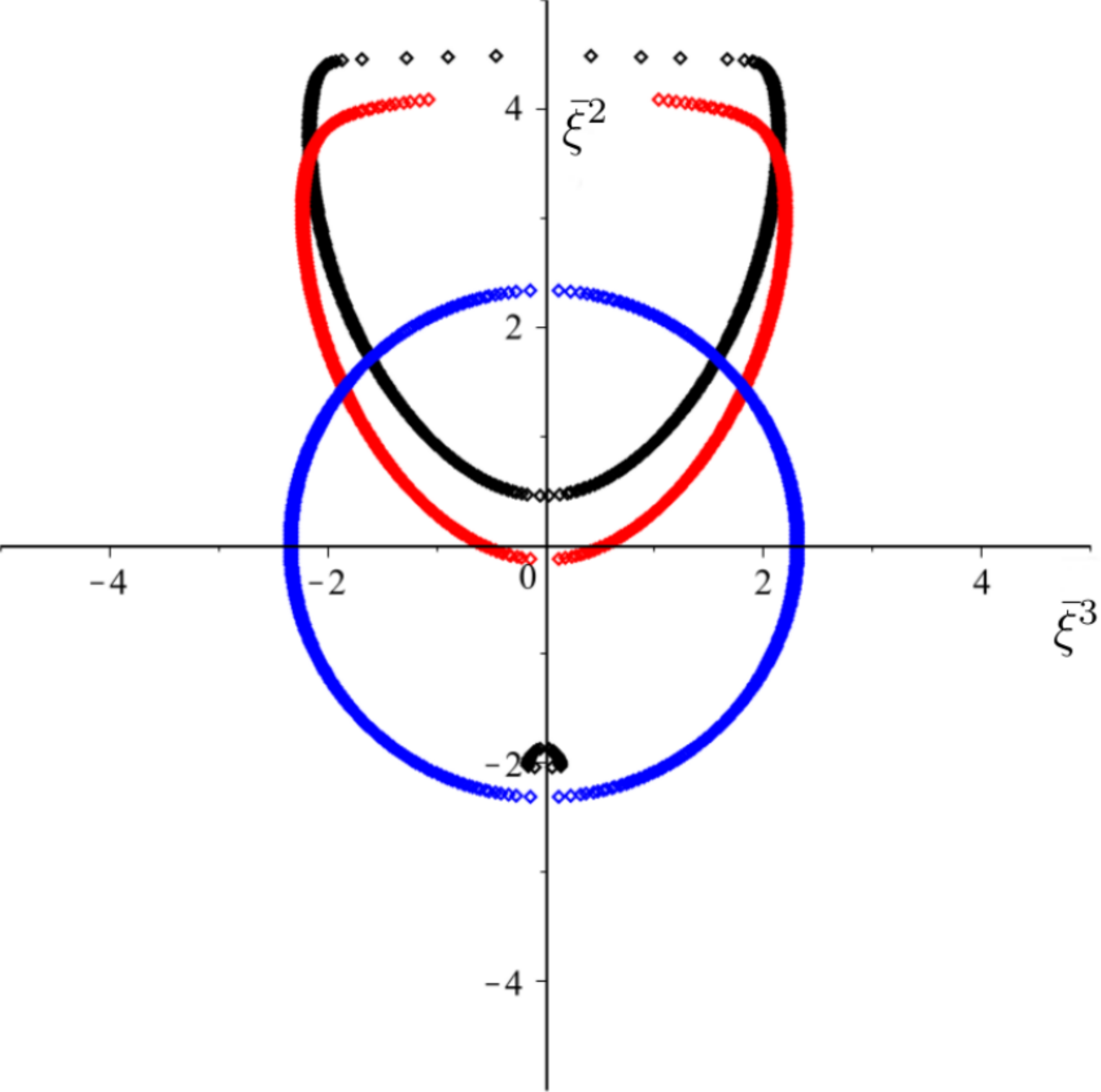}
  \ea
  \caption{ \label{Fig_7} On the top we can see the shadow of a black hole for an observer at $\theta_o=\pi/4$ and radius $r_o=5$; the values of the multiple moments are $c_1=\frac{1}{800}$, (with red, (3)) $c_1=\frac{1}{1000}$, (with black, (2)) and $c_1=0$, (which is the undistorted case with blue, (1)). On the bottom we can see the shadow of a black hole for an observer at $\theta_o=\pi/2$ and radius $r_o=5$; the values of the multiple moments are $c_1=\frac{1}{1000}$, (with red, (2)) $c_1=\frac{1}{800}$, (with black, (3)) and $c_1=0$, (which is the undistorted case with blue, (1)).}
\end{center}  
\end{figure}
\begin{figure}[htb]
\begin{center}
\ba
&&\includegraphics[width=6cm]{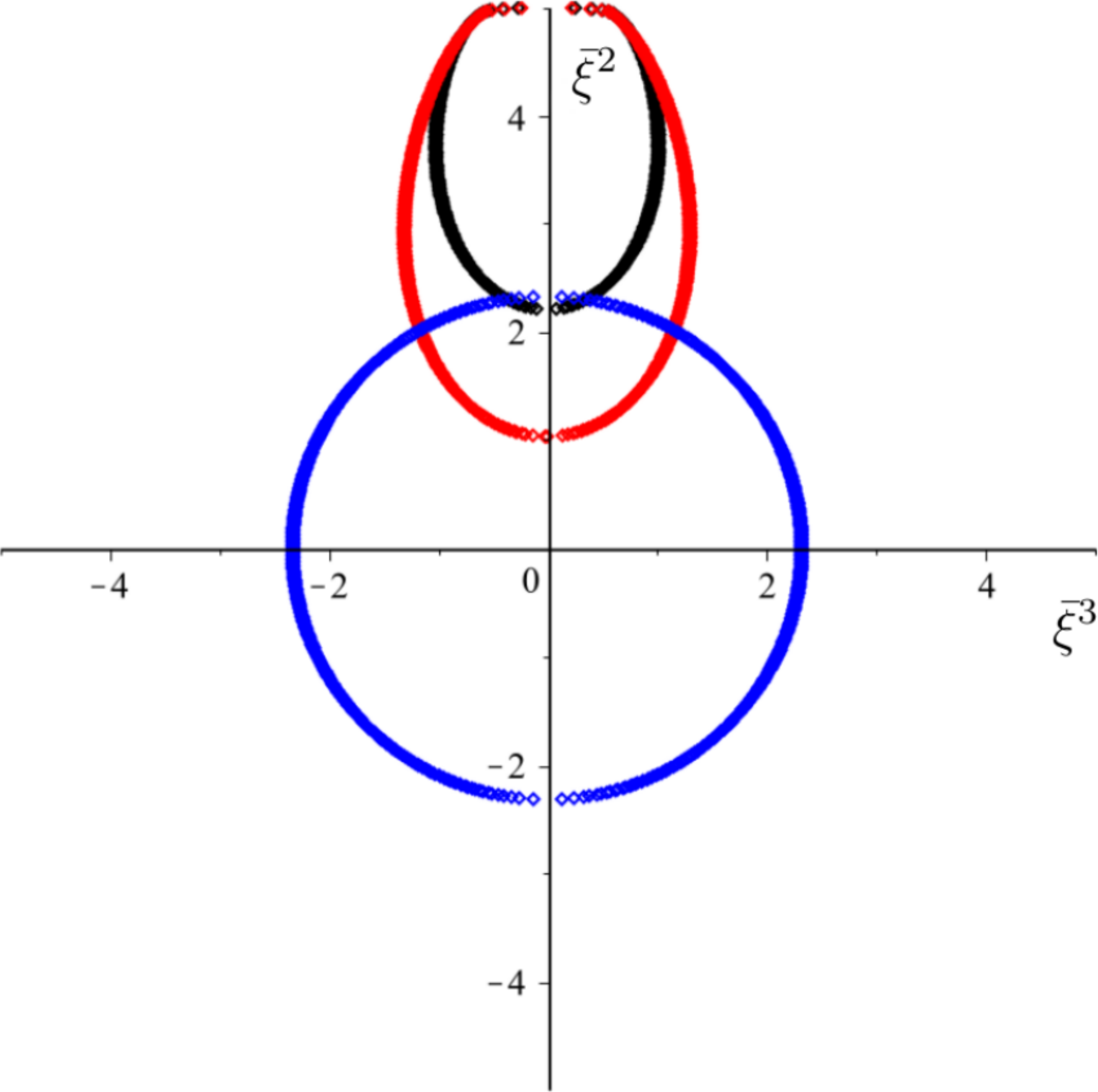}\nonumber\\
&&\includegraphics[width=6cm]{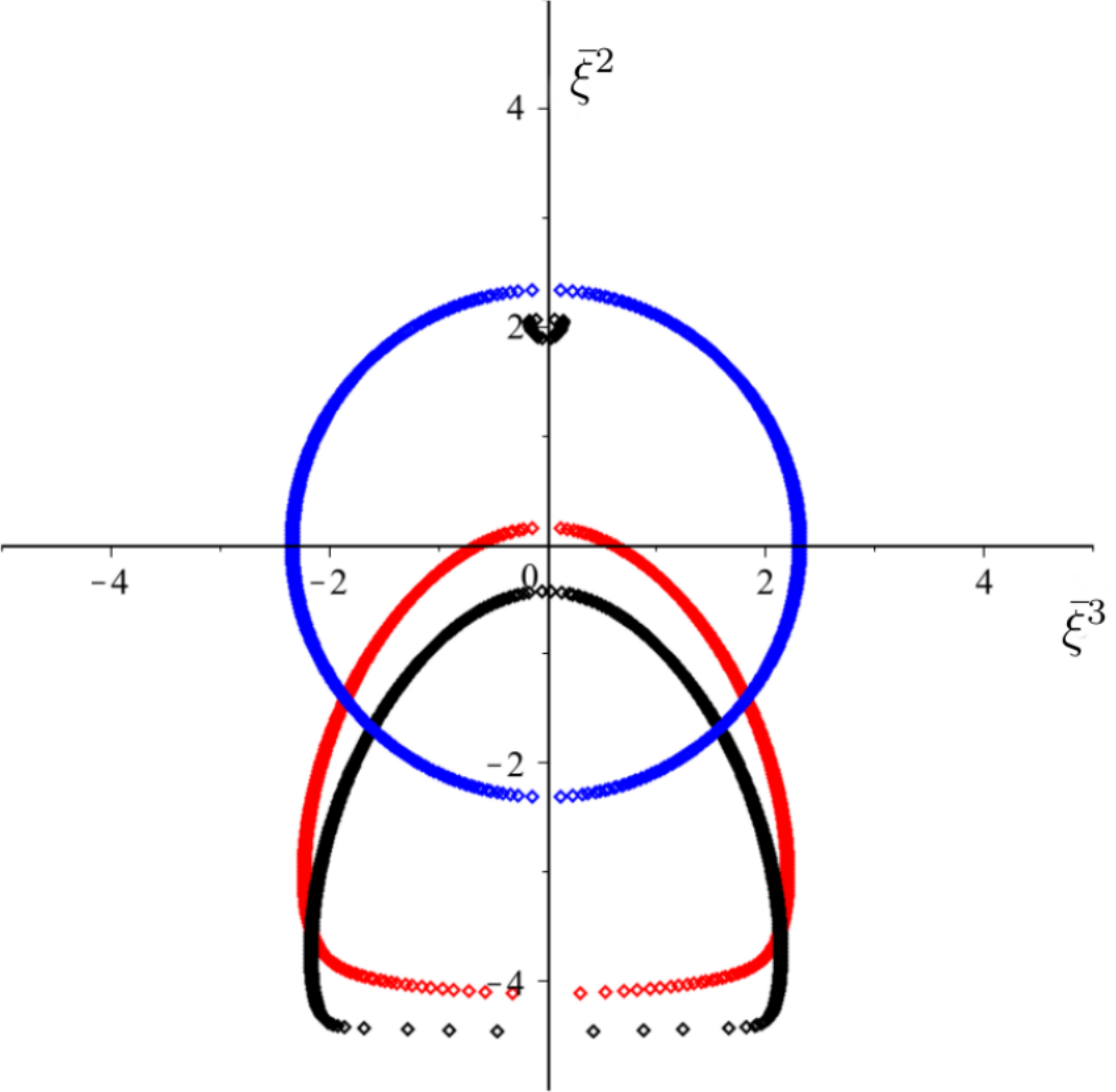}
  \ea
  \caption{ \label{Fig_8} On the top we can see the shadow of a black hole for an observer at $\theta_o=\pi/4$ and radius $r_o=5$; the values of the multiple moments are $c_1=-\frac{1}{1000}$, (with red, (2)) $c_1=-\frac{1}{800}$, (with black, (3)) and $c_1=0$, (which is the undistorted case with blue, (1)). On the bottom we can see the shadow of a black hole for an observer at $\theta_o=\pi/2$ and radius $r_o=5$; the values of the multiple moments are $c_1=-\frac{1}{800}$, (with red, (3)) $c_1=-\frac{1}{1000}$, (with black, (2)) and $c_1=0$, (which is the undistorted case with blue, (1)).}
\end{center}  
\end{figure}

We present the effect of distortions for even multiple moments in figures \ref{Fig_5}-\ref{Fig_6}. We see that for an observer located in the equatorial plane, $\theta_o=\pi/2$ the shadow looks like an ellipse. For positive and negative even multiple moments $c_2$, we respectively obtain prolate and oblate shadow shapes. As the  multiple moment increases through positive values, the shadow becomes more prolate, whereas it becomes more oblate for increasingly negative values. For an observer not located on the equatorial plane, the local shadow of the black hole moves upward or downward in the impact plane of observer. As the magnitude of the negative multiple moment increases, the shadow  moves down (when $\bar{\xi}^2$ is the horizontal axes, illustrated in figure \ref{Fig_5}). As the value of positive multiple moment increases the shadow moves up, as shown in figure \ref{Fig_6} . 

\begin{figure}[htb]
\begin{center}
\ba
&& \includegraphics[width=6cm]{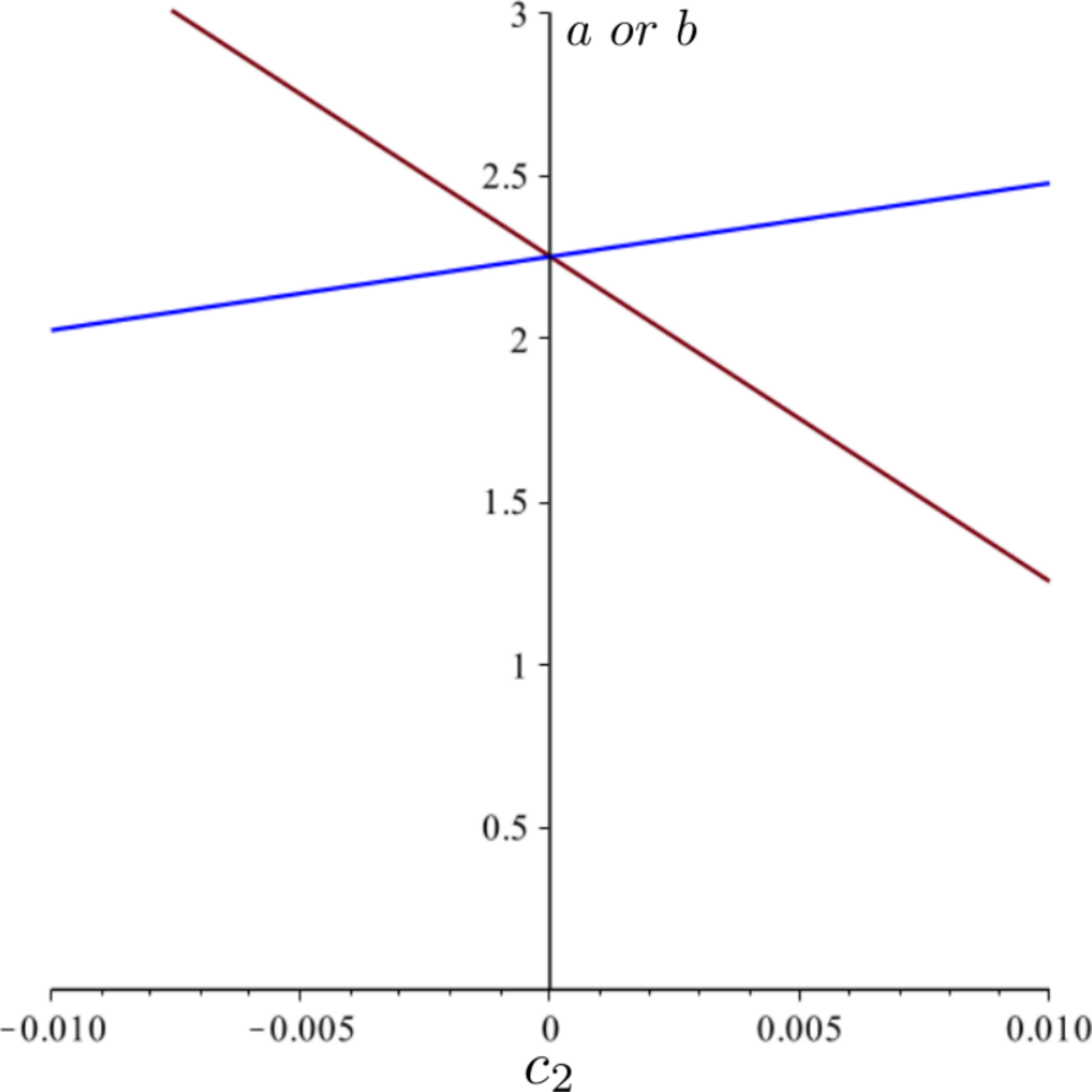}\nonumber\\
&& \includegraphics[width=6cm]{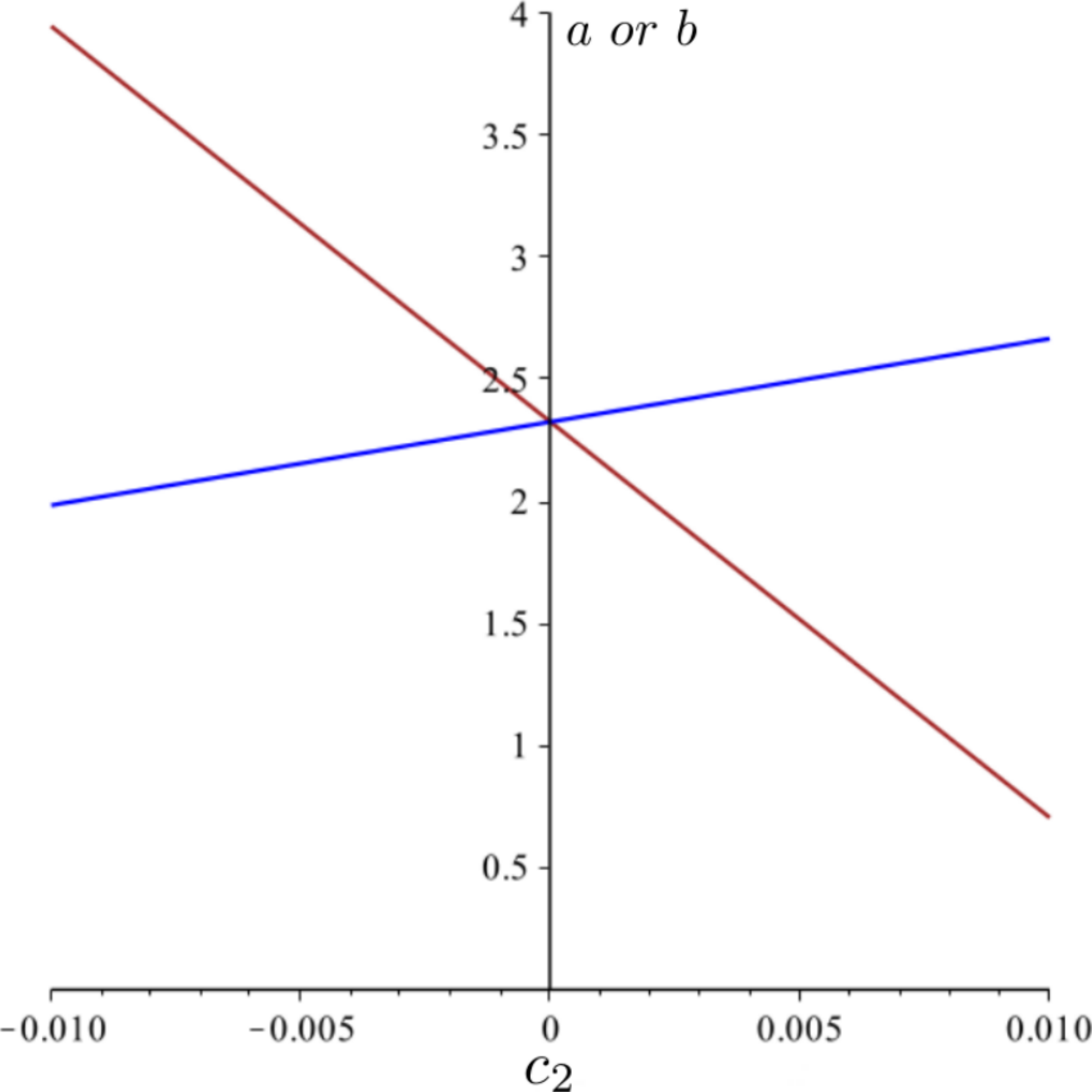}
\ea
  \caption{ \label{Fig_9} The parameters of the ellipse $a$ (red, (2)) and $b$ (blue, (1)) vs.  the multiple moment $c_2$. Top: observer at $r_o=4$; Bottom: observer   at $r_o=5$. }
\end{center}  
\end{figure}

Since the shadow looks like an ellipse for an observer located at $\theta_o=\pi/2$, we fit the shape of the shadow to an ellipse, taking $a$ and $b$ to be its  semi-major/semi-minor axes, the origin at the centre,  
and the  angular coordinate $\phi$ to be measured from the axis $\bar{\xi}^3$. To  find $a$ and $b$ we minimize the normalized square error 
\be
J=\frac{\sum_{i=0}^m (\rho^{\text{numer}}_i-\rho_i)^2}{\sum_{i=0}^m(\rho^{\text{numer}}_i)^2},\n{fit1}
\ee
with
\ba
&&\rho^{\text{numer}}_i=[(\bar{\xi}^{3})^2+(\bar{\xi}^{2})^2]^{1/2},\nonumber\\
&&\rho_i=\frac{ab}{\sqrt{b^2\cos^2\phi_i+a^2\sin^2\phi_i}}, \nonumber\\
&&\phi_i=\arctan(\frac{\bar{\xi}^{2}}{\bar{\xi}^{3}}),\n{fit2}
\ea
where $\rho^{\text{numer}}_i$ is given by our numerical result, and $\rho_i$ is the equation of the ellipse for every $\phi_i$. 
In figure \ref{Fig_9} we plot the parameters $a$ and $b$ as a function of the quadrupole moment $c_2$ for an observer located on the equator ($\theta_o=\pi/2$) and at a fixed distance $r_o$ from the centre of the black hole. To find this relationship, we assumed that $a$ and $b$ are linear functions of $c_2$ and that the intercept is given by the value $({27(r_o-1)}/{4r_o})^{1/2}$ (see equation (\ref{M1})). Obtaining  the slope,  we see that the assumption of linearity between $a$ or $b$ and $c_2$ is a good one, since it gives a fitting error comparable or smaller than the size of our pixel, (recall that this is $0.01$).  
For the observer on the equatorial plane and distance $r_o=4$, we find  
\be
a=-99.68c_2+2.25,~~~b=22.58c_2+2.25,
\ee 
where the respective errors in the fits for $(a,b)$ are $(4.3\times 10^{-4}, 1.1\times 10^{-6})$. As stated, these errors are smaller than the size $0.01$ of our pixel, and so this is the actual error. For $c_2=0$ the shadow is a circle of radius $2.25$.
For an observer on the equatorial plane at  $r_o=5$, the values  are  
\be
a=-161.96c_2+2.32,~~~b=34.11c_2+2.32,
\ee 
where the fitting errors for $(a,b)$ are now $(3.5\times 10^{-2},1.2\times 10^{-4})$ respectively. For $c_2=0$ the shadow is a circle of radius $2.32$. As the observer moves away from the black hole the radius of the circle increases. For the undistorted Schwarzschild geometry ($c_2=0$), and  the shadow is a circle of radius $3\sqrt{3}/2\sim 2.598$ (see \eq{ShwShadow0}) for
an observer located at infinity.  Table \ref{Table1} indicates that as the observer is moves away from the black hole the magnitudes of the slopes of the parameters $a$ and $b$ increase. This is also valid for the intercepts. 
  \begin{table*}[t]
\caption{\label{Table1} Obtained values and errors of the parameters $a$ and $b$ for an observer on the equatorial plane at $r_o$.}
\begin{tabular}{| c| c| c | c| c | }
\hline
 $r_o$ & $a$ & error in $a$ fitting & $b$ & error in $b$ fitting\\
 \hline
    3 & $a=-41.85c_2+2.12$ & $1.34\times 10^{-4}$ & $b=8.89c_2+2.12$ & $7.25\times 10^{-4}$\\
    4  & $a=-99.68c_2+2.25$&  $4.3\times 10^{-4}$ & $b=22.58c_2+2.25$ &$1.1\times 10^{-6}$ \\
    5 & $a=-161.96c_2+2.32$ & $3.5\times 10^{-2}$ & $b=34.11c_2+2.32$ & $1.2\times 10^{-4}$\\
    6 & $a=-251.63c_2+2.37$ & $2.4\times 10^{-2}$ & $b=44.27c_2+2.37$ & $3.3\times 10^{-5}$\\
   \hline
\end{tabular}
\end{table*}  

Though it seems that the parameters $a$ and $b$ vanish for large values of $c_2$,
we emphasize that these slopes are valid only for small values  $-1/150<c_2<1/150$ of the quadrupole moment. Had we 
included larger values, say $-1/5<c_2<1/5$, then the slopes in   Table \ref{Table1} would have been totally different, since  distortions drastically dominate over the black hole potential for larger values of the multiple moments.

The comparison of the shadow of a distorted black hole to the embedding of its horizon is interesting. The horizon surface, $t=\text{const}$, $r=1$, of the distorted black hole \eq{ST4} is given by 
\ba
dS_H^2=e^{-2\mathcal{U}_H} (e^{2 {V}_H}d\theta^2+\sin^2{\theta}d\phi^2) \n{HorizonMetric}
\ea
where, $\mathcal{U}_H =\mathcal{U}(r=1,\theta)$ and $V_H =V(r=1,\theta)$.
Consider   isometrically embedding   this 2-dimensional axisymmetric metric into the 3-dimensional space
\ba
ds^2=\epsilon dz^2+d\rho^2+\rho^2 \n{Embed}
\ea
where $(Z,\rho,\phi)$ are   cylindrical coordinates. Setting $\epsilon=1$ corresponds to Euclidean space, whereas $\epsilon=-1$ corresponds to pseudo-Euclidean space. We have $Z=Z(\theta)$, and $\rho=\rho(\theta)$. Matching the metrics ($\ref{HorizonMetric}$) and ($\ref{Embed}$), we derive the following embedding map
\ba
&&\rho(\theta)=e^{-{\mathcal{U}_H}}\sin\theta, \nonumber\\
&&Z(\theta)=\int_{\theta}^{\pi/2}\left[\epsilon(e^{-2\mathcal{U}_H+2V_H}-\rho_{,\theta'}^2)\right]^{\frac{1}{2}}d\theta'.
\ea
For even multiple moments  the horizon is deformed to an ellipse. Using the same fitting method   as before, we find
\be
a=1.07c_2+1,~~~ b=-2.11 c_2+1 \n{Hembedding}
\ee 
where we have fixed the intercept of the linear fit to be $1$ and found the slope; the errors in the above fittings are of order $10^{-5}$.

Quite counterintuitively, for positive $c_2$ the horizon of the black hole is oblate whilst its shadow is prolate. For the negative $c_2$ the opposite occurs: a prolate horizon has an  oblate shadow. As noted above, a positive $c_2$  corresponds to a ring on the equatorial plane in Newtonian gravity. We can see from the bottom plot of figure \ref{com} that for  $c_2 > 0$, $g_{ttd}$ of the distorted black hole around the equator is less than that of an undistorted Schwarzschild black hole, whereas on axis $g_{ttd}$ is greater. This implies that the gravitational field on the equator for a distorted black hole is greater than the undistorted case and weaker on axis\footnote{Note that for $c_2>0$ the function  \[\mathcal{U}=-\frac{c_2}{2}-\frac{c_2}{2}\left(2r-1\right)^2 \, , \] 
on the equator is always negative. Therefore $f=g_{ttd}/g_{tt}<1$.}. On the equator there exists the effect of the gravitational field of both the ring and the black hole. For a quadrupole  moment $c_2$ the equations of motion imply that on the equatorial plane  $\ddot{\theta}=0$ when 
$\dot{\theta}=0$ (note that $f_{,\theta}$ and $h_{,\theta}$ vanish on the equator in this case). This implies the existence of planar motion. Since the equations of motion are integrable on the equator, we can construct an effective potential $V$ for the planar motion of null rays \cite{Wheeler}
\begin{figure}[htb]
\begin{center}
\hfill
  \includegraphics[width=6cm]{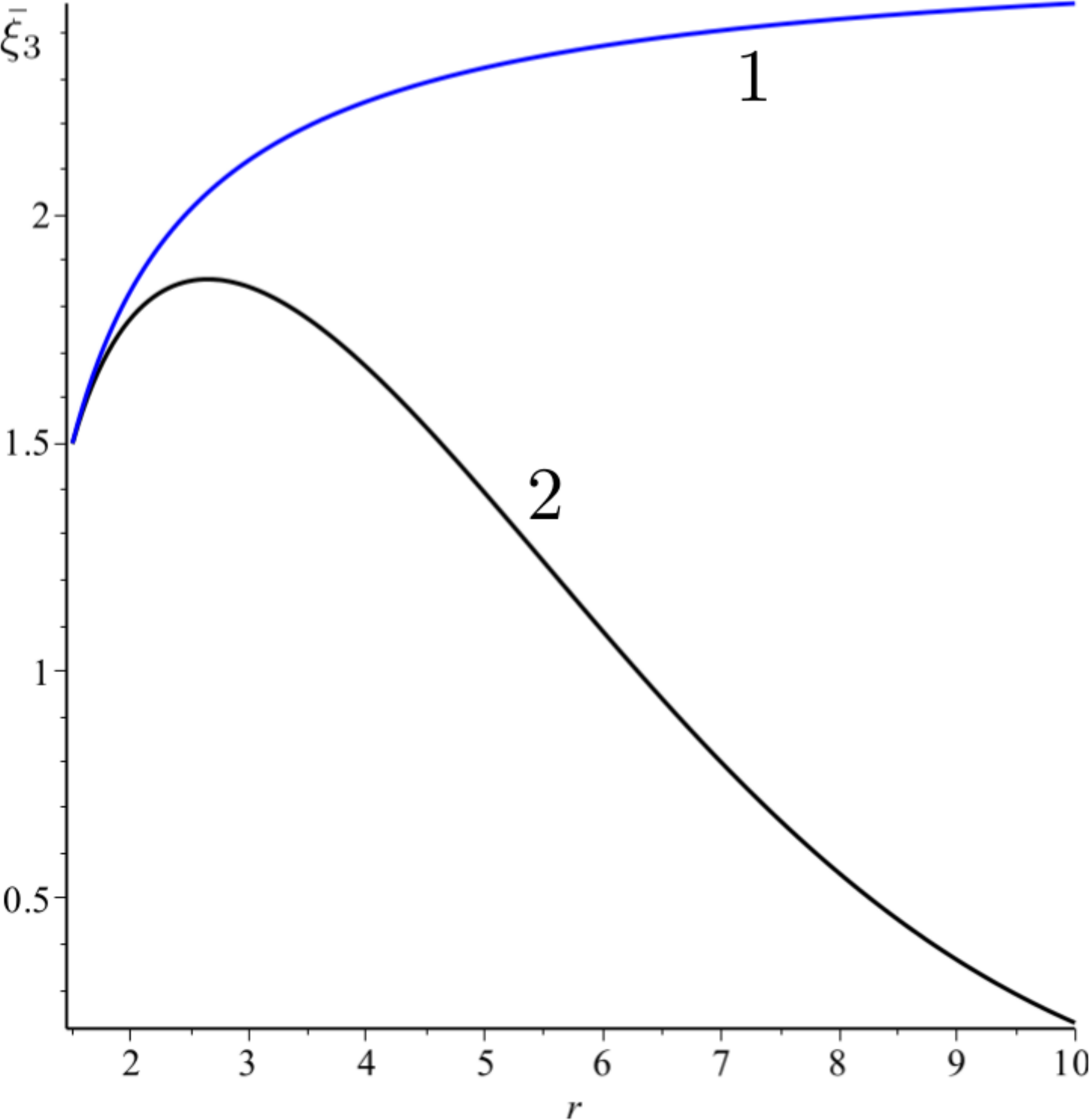}
  \hfill  \, { }
  \caption{ $\bar{\xi}_3$ as a function of the position of the observer in the equatorial plane.The blue, (1), colour corresponds to the undistorted case whereas the black, (2), colour corresponds to  the distorted one. The value of $c_2=1/150$. \label{xi3com}  
  \label{xi3com}}
\end{center} 
\end{figure}

\ba
\frac{1}{l_z^2}(\frac{d\tilde{r}}{d\tilde{t}})^2=\frac{1}{l_z^2}-V^2,
\ea 
where
\ba
&&V^2=\frac{1}{r^2}(1-\frac{1}{r})f^2, \nonumber\\
 &&d\tilde{t}=[f(1-\frac{1}{r})]^{1/2},\nonumber\\
&& d\tilde{r}=[h(1-\frac{1}{r})^{-1}]^{1/2}  \,  .
\ea 
The maximum of the effective potential is equal to the value of $1/l_z^2$ or $1/{b_{critical}^2}$ for the knife-edge orbit. For $c_2=1/150$ we have $1/{b_{critical}^2}=0.1389$. Using \eq{t2a}, for an observer located on the equator and $r_o=5$ this corresponds to $\bar{\xi}^3=1.3894$, which shows that the shadow of the distorted black hole on the equator has a smaller diameter than that of an undistorted Schwarzschild black hole, for which  $\bar{\xi}^3=2.32$  confirming the result of a prolate shadow for $c_2> 0$. The knife-edge orbit is located at radius $r=1.50$, (or, $\rho=3M$), for the undistorted case. For  $c_2=1/150$ this orbit is at $r=1.46$. Evidently the attractive influence of the ring allows potentially escaping null rays to more closely approach the black hole. 

The angular momentum of the photons for the knife edge orbit is larger than the undistorted case. If instead of \eq{t2a}, which is valid for a local observer, we had the map \eq{t2} (which is for an observer at infinity) this would immediately imply that $\bar{\xi}_3$ would be larger for a distorted black hole. However the function $f$ plays an important role in \eq{t2a}; on the equator it is always less than one and decreases as $r$ increases. As a consequence $\bar{\xi}_3$ decreases with distance, implying the horizontal axis of the ellipse decreases in size as the observer gets further and further away from the black hole. For an observer very close to the black hole $\bar{\xi}_3$ has almost the same value of that of a Schwarzschild
black hole. This is illustrated in figure \ref{xi3com}. 

While our metric represents a vacuum space-time, there must be (more distant) matter sources  causing distortion of
 the black hole. Provided these  sources are located in some finite-sized region
we can extend the solution beyond this (non-vacuum) region to a yet more distant vacuum region. In this exterior region we must replace the expansions of the functions $\mathcal{U}$ and $V$ with expansions in terms of exterior multiple moments. In this exterior vacuum region the function $f$  increases until becomes unity at infinity. Assuming the angular momentum of the photon is conserved as it (traced backward) passes through the non-vacuum and exterior vacuum regions,  an observer in the exterior vacuum region can use the critical value of the angular momentum of the photon to deduce $\bar{\xi}_3$. For an observer located at infinity $\bar{\xi}_3$ is larger than that of the undistorted Schwarzschild black hole. 

In figure \ref{LF1} we illustrate a schematic representation of an oblate horizon and a prolate shadow. All the observers in the $(x,y)$ plane (at any $\phi$ angle) see the same  shadow shape. For illustration purposes the projection of the shadow was made on the other side of the black hole rather than on the plane of the black hole.
\begin{figure}[htb]
\begin{center}\label{LF1}
 \includegraphics[width=8cm]{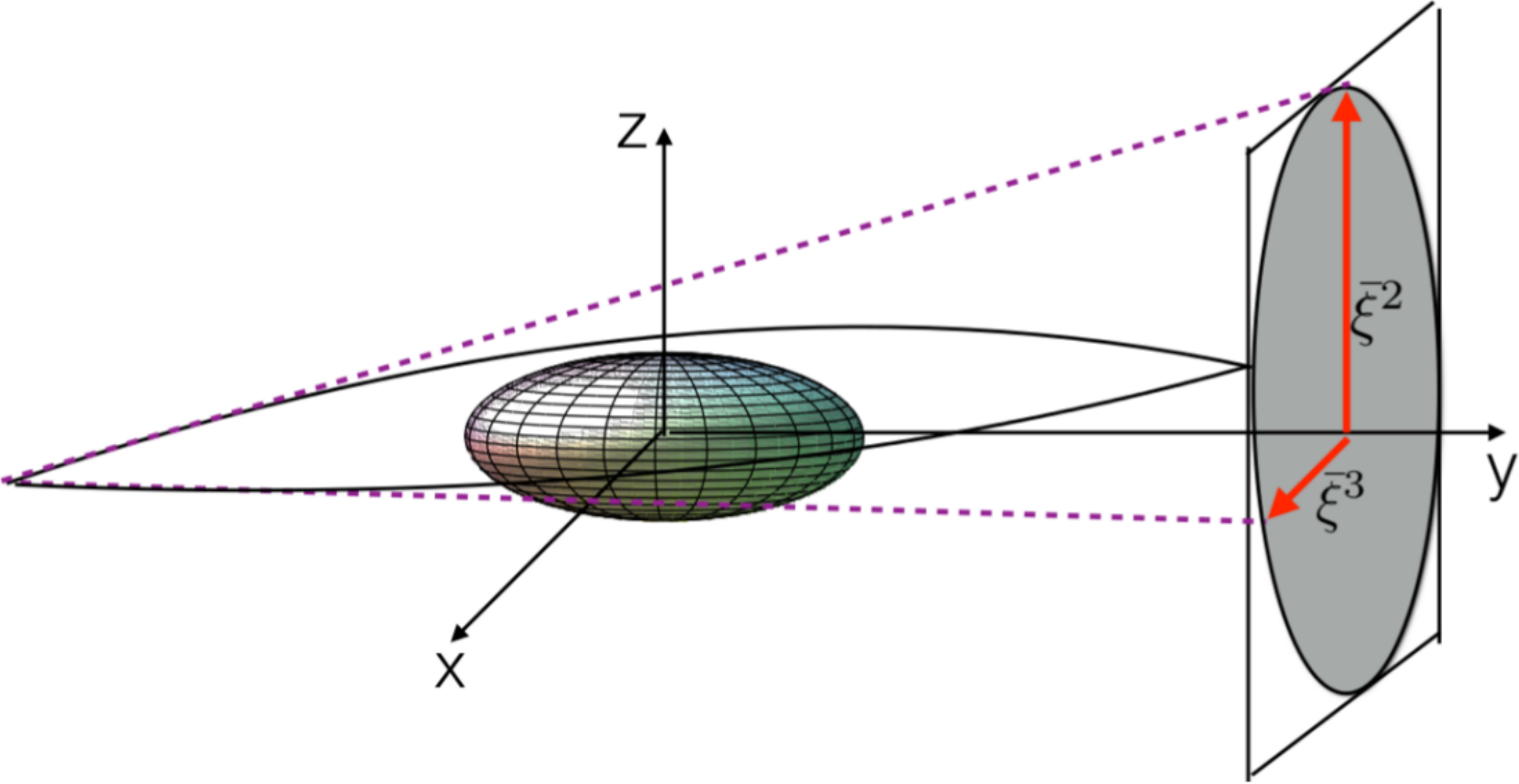}
   \caption{ Schematic representation of an oblate horizon and a prolate shadow. }\label{LF1}
\end{center}  
\end{figure}\label{LF1}
One can say that the black hole horizon is more ``rigid" than that of the black hole shadow, i.e. a larger value of the multipole moment $c_2$ is required in order to deform the horizon as much as the black hole shadow.
 
Note that for very large values of multiple moments (which we do not consider), the horizon cannot be embedded in Euclidian space. For a 2-dimensional axisymmetric metric if the Gauss curvature is negative at the fixed points of the rotation group, it is impossible to isometrically embed a region containing such a fixed point in Euclidean space $\mathbb{E}^{3}$. Such surfaces can be globally embedded in $\mathbb{E}^{4}$ \cite{Berger}. In the case of Schwarzschild black hole, with quadrupole distortion the Gauss curvature for $c_{2}>1/12$ becomes negative at both of the poles, $\theta=0$, and $\theta=\pi$. Similarly, for the octupole distortion, the Gauss curvature becomes negative at one of the poles for $|c_{3}|>1/20$.  For multipole moments greater than or equal to these values, the horizon surface of the distorted black hole cannot be isometrically embedded in a flat 3D space.

For odd multiple moment $c_1$, the shadow is illustrated in figures \ref{Fig_7}-\ref{Fig_8}. Here we observe an ``eyebrow'' structure for large values of multiple moments (e.g. $c_1=1/800$), reminiscent of those seen for the shadow of two merging black holes \cite{merge}. For an observer in the equatorial plane the shadow looks like an incomplete ellipse that  is flattened on one side; furthermore   the shape of the shadow is mirror-reflected with respect to the $\bar{\xi}^3$ axis upon changing the sign of the multipole moment from positive to negative. For an observer in the $\theta_o=\pi/4$ plane, we see the negative odd multiple moment to have a shape similar  to that of the positive even multiple moment in the same plane; however the width of the shapes are smaller than the ones for even multiple moment.

\section{Summary}

 In many astrophysical situations, such as a black hole in a binary system, the black hole is distorted. Light rays emitted from sources behind the black hole will either be absorbed by it or escape to an observer located at some finite distance
$(r_{o},\theta_{o})$ from the black hole.  If a photon gets emitted from a point located on a sphere of radius ${r}_{e}$,
the boundary of the set of rays reaching this observer bounds the  `local shadow' of the black hole for this observer. 

 We have introduced this notion of local shadow in precise terms and computed it for a distorted Schwarzschild black hole by
 tracing back the trajectory of these photons. A null ray traced back from the point $(r_{o},\theta_{o})$, with initial parameters $(\bar{\xi}^{2}, \bar{\xi}^{3})$ will be part of the local shadow if this   null ray is absorbed by the black hole; if it  reaches a radius $r_{e}$ (after propagating in the space-time) it is not part of  ``local shadow'' of the black hole. The parameters $(\bar{\xi}^{2}, \bar{\xi}^{3})$ are related to the initial velocity $\theta'$, integral of motion $l_z$, and the position of the observer. 

 For a quadrupole distortion, $c_2\neq 0$ we found that the shadow was deformed from a circle to an ellipse. As the magnitude of the quadrupole moment $c_2$ increases, the shadow gets increasingly  deformed . Rather unexpectedly, we found that
  positive values of $c_2$ the shadow is prolate whereas the horizon is oblate; for $c_2<0$ the reverse occurs, with a prolate horizon having an oblate shadow. We have shown analytically why this phenomenon occurs for a local observer. However we expect that an observer at infinity will see the apparent shape of an oblate black hole (i.e., its shadow) to be either oblate or a circle of greater radius than  for the undistorted case\footnote{This oblate/circle  ambiguity is due to the fact that we cannot deduce the value of $\bar{\xi}^{2}$ at infinity analytically.}. The horizon is very rigid, i.e., it is harder to deform the horizon and easier to have a deformed shadow. In this paper, we don't consider large enough distortions which distort the horizon effectively, since we do not want the distorting potentials to dominate over the potential of the black hole.

\begin{acknowledgments}
The authors are grateful to the Natural Sciences and Engineering Research Council of Canada for financial support. The authors S. A. and C. T. also acknowledge support by the DFG Research Training Group 1620 ``Models of Gravity''. The authors would like to acknowledge Don N. Page for a useful comment. 
\end{acknowledgments}

\end{document}